\documentclass{aa}

\usepackage[pdftex,pdfpagemode={UseOutlines},bookmarks,bookmarksopen,colorlinks,linkcolor={blue},citecolor={blue},urlcolor={red}]{hyperref}
\usepackage{graphicx}
\usepackage{subfigure}
\usepackage{subfloat}
\usepackage{amsmath}
\usepackage{txfonts} 
\usepackage{microtype}
\usepackage{tabularx}
\usepackage{array}
\usepackage{adjustbox}
\usepackage{balance}
\usepackage[utf8]{inputenc} 

\begin{document} 
\title{The EBLM Project}
\subtitle{V. Physical properties of ten fully convective, very-low-mass stars}

\author{Alexander von Boetticher\inst{1,2}
	\and Amaury H.M.J. Triaud\inst{3,4}
	\and Didier Queloz\inst{2,5}
    \and Sam Gill\inst{6}
	\and Pierre F.L. Maxted\inst{6}
    \and Yaseen Almleaky \inst{7}
	\and David R. Anderson\inst{6}
    \and Francois Bouchy\inst{5}
	\and Artem Burdanov \inst{8}
	\and Andrew Collier Cameron\inst{9}
	\and Laetitia Delrez\inst{2}
	\and Elsa Ducrot \inst{8}
	\and Francesca Faedi\inst{10}
	\and Micha\"el Gillon\inst{8}
	\and Yilen G\'omez Maqueo Chew\inst{11}
	\and Leslie Hebb\inst{12}
	\and Coel Hellier\inst{6}
	\and Emmanu\"el Jehin\inst{8}
    \and Monika Lendl\inst{13,14}
	\and Maxime Marmier \inst{5}
	\and David V. Martin\inst{5, 15}
	\and James McCormac \inst{10}
	\and Francesco Pepe \inst{5}
	\and Don Pollacco\inst{10}
	\and Damien S\'egransan \inst{5}
	\and Barry Smalley\inst{6}
	\and Samantha Thompson\inst{2}
	\and Oliver Turner \inst{5}
	\and St\'ephane Udry \inst{5}
	\and Valérie Van Grootel \inst{8}
	\and Richard West\inst{10}
}

\offprints{alexander.vonboetticher@physics.ox.ac.uk}

\institute{Clarendon Laboratory, Parks Road, Oxford, OX1 3PU, UK
	\and Cavendish Laboratory, J J Thomson Avenue, Cambridge, CB3 0HE, UK
	\and School of Physics \& Astronomy, University of Birmingham, Edgabston, Birmingham B15 2TT, UK 
	\and Institute of Astronomy, Madingley Road, Cambridge CB3 0HA, UK
	\and Observatoire Astronomique de l'Universit\'e de Gen\`eve, Chemin des Maillettes 51, CH-1290 Sauverny, Switzerland
    \and Astrophysics Group, Keele University, Staffordshire, ST5 5BG, UK
    \and King Abdullah Centre for Crescent Observations and Astronomy, Makkah Clock, Mecca 24231, Saudi Arabia
    \and Universit\'e de Li\`ege, All\'ee du 6 ao\^ut 17, Sart Tilman, 4000, Li\`ege 1, Belgium
    \and SUPA, School of Physics \& Astronomy, University of St Andrews, North Haugh, KY16 9SS, St Andrews, Fife, Scotland, UK
    \and Department of Physics, University of Warwick, Coventry CV4 7AL, UK
    \and Instituto de Astronom\'ia, Universidad Nacional Aut\'onoma de M\'exico, Ciudad Universitaria, Ciudad de M\'exico, 04510, M\'exico
    \and Hobart and William Smith Colleges, Department of Physics, Geneva, NY 14456, USA
	\and Space Research Institute, Austrian Academy of Sciences, Schmiedlstr. 6, 8042, Graz, Austria
	\and Max Planck Institute for Astronomy, K\"onigstuhl 17, 69117 Heidelberg, Germany
	\and University of Chicago, 5640 S Ellis Ave, Chicago, IL 60637, USA
}


\abstract{Measurements of the physical properties of stars at the lower end of the main sequence are scarce. In this context we report masses, radii and surface gravities of ten very-low-mass stars in eclipsing binary systems, with orbital periods of the order of several days. The objects probe the stellar mass-radius relation in the fully convective regime, $M_\star \lesssim 0.35$ M$_\odot$, down to the hydrogen burning mass-limit, $M_{\mathrm{HB}} \sim 0.07$ M$_\odot$. The stars were detected by the WASP survey for transiting extra-solar planets, as low-mass, eclipsing companions orbiting more massive, F- and G-type host stars. We use eclipse observations of the host stars, performed with the TRAPPIST, \textit{Leonhard Euler} and SPECULOOS telescopes, and radial velocities of the host stars obtained with the CORALIE spectrograph, to determine the physical properties of the low-mass companions. Surface gravities of the low-mass companions are  derived from the eclipse and orbital parameters of each system. Spectroscopic measurements of the host star effective temperature and metallicity are used to infer the host star mass and age from stellar evolution models for solar-type stars. Masses and radii of the low-mass companions are then derived from the eclipse and orbital parameters of each binary system.  
The objects are compared to stellar evolution models for low-mass stars by calculating residuals with respect to model predictions. The residuals are used to test for an effect of the stellar metallicity and orbital period on the radius of low-mass stars in close binary systems. Measurements are found to be in good agreement with stellar evolution models; a systematic inflation of the radius of low-mass stars with respect to model predictions is limited to 1.6 $\pm$ 1.2\%, in the fully convective low-mass regime. The sample of ten objects indicates a scaling of the radius of low-mass stars with the host star metallicity. No correlation between stellar radii and the orbital periods of the binary systems is determined. A combined analysis with thirteen comparable objects from the literature is consistent with this result. 
\bigskip 
}
\maketitle

\section{Introduction}

Low-mass stars are the most common stellar objects in our galaxy \citep{Kroupa_2001, Chabrier_2003, Henry_2006} and are ubiquitous in the solar neighbourhood. At least $\sim$70\% of stars within 10 pc of the sun are of spectral type M \citep{Henry_2006}, corresponding to a mass $\lesssim 0.6$ M$_\odot$. Such low-mass stars are excellent candidates for the detection and characterisation of Earth-sized extra-solar planets, since the photometric signal of a transiting planet scales inversely with the square of the host star size. Orbits of temperate stellar irradiation around such stars have periods of the order of days, increasing the geometric probability of a transit, and enabling frequent observations of transit events. The system of seven Earth-sized exoplanets orbiting the very-low-mass star TRAPPIST-1 \citep{Gillon_2016,Gillon_2017} demonstrates the significance of very-low-mass stars to exoplanet science.
An understanding of the properties of very-low-mass stars is required to reliably characterize the planets that orbit such stars, since the derivation of planetary parameters is usually dependent on the parameters of their stellar hosts. The mass and radius of very-low-mass stars are fundamental in this respect, but are difficult to determine for field stars, due to their low brightness \citep[e.g. $M_V \sim 18.4$ for TRAPPIST-1, with $M_\star = 0.089 \pm 0.006$ M$_\odot$;][]{vanGrootel_2018}. The stellar mass and radius can however be readily measured for components in eclipsing binary systems. Such measurements have revealed discrepancies between empirical radii and predictions by stellar evolution models: the radii of low-mass stars are frequently inflated with respect to stellar models, by up to 20\% \citep[eg.][]{Hoxie_1973, Lacy_1977, Morales_Ribas_2005, Torres_2010, Spada_2013}. The inflated radii of the well-characterized eclipsing binaries CM Draconis \citep[e.g.][]{Morales_2009} and YY Geminorum \citep{Torres_Ribas_2002} are archetypal. Discrepancies in temperature have also been reported \citep{ofir2012,gomez2014}. A recent study on white-dwarf + M dwarf binaries by \citet{Parsons_2018} indicates a $5\%$ systematic bias towards larger radii for a sample of fully and partially convective low-mass stars, with good agreement between effective temperatures and model predictions.
Proposed mechanisms to explain the inflation of radii of low-mass stars include metallicity effects \citep{Berger_2006}, and stellar magnetic activity \citep{Mullan_MacDonald_2001, Chabrier_2007}. The activity hypothesis is supported by observations that indicate that inflated radii are especially prevalent for low-mass stars in short period binary systems \citep[e.g.][]{Morales_2009, Kraus_2011, Spada_2013}. The tidal interaction in such systems can give rise to a fast rotation of the components, that is expected to generate strong magnetic fields via a dynamo mechanism. It has been suggested that the rotation itself, and the associated magnetic activity \citep{Gough_1966, Chabrier_2007} can inhibit the convective transport of heat in low-mass stars, leading to larger stellar radii. 

The occurrence of stars with inflated radii in the mass-regime $M_\star \gtrsim 0.35$ M$_\odot$ is well established by observations \citep[eg.][]{Spada_2013}. In the range $M_\star \lesssim 0.35$ M$_\odot$, fewer stars with mass and radius measurements are available, due to the low brightness of such cool and small objects. In this paper, we examine this very-low-mass regime, that we identify with $M_\star \lesssim 0.35$ M$_\odot$. The distinction between the two mass ranges is motivated by the expected structural change of the stellar interior at $M_\star = 0.3 - 0.4$ M$_\odot$: stellar models indicate that stars with masses $\gtrsim 0.35$ M$_\odot$ have radiative cores, whereas objects with lower masses are fully convective throughout the stellar interior \citep{ChabrierBaraffe_1997}. This difference in the interior physics has implications for the generation of magnetic fields via a dynamo mechanism. In fully convective stars, that lack a tachocline, the shell-dynamo of stars with a radiative core cannot operate. Magnetic fields can arise in a different manner; the distributive dynamo, driven by convective turbulence is a favoured candidate \citep[eg.][]{Mullan_MacDonald_2001}. \citet{Mullan_MacDonald_2001} provide a detailed discussion of the magnetic properties of low-mass stars and of the transition between radiative and completely convective interiors; \citet{Browning_2008} have performed simulations of magnetic dynamo activity in fully convective stars. Empirical measurements are required to determine if the different interior and dynamo physics of fully convective stars manifests in observational properties that are distinct from those of more massive low-mass stars that possess a radiative core.

We report eclipse observations and spectroscopic measurements of ten binary systems with low-mass components in the fully convective mass range. Each system consists of an F- or G-type, bright primary star, with a very-low-mass companion that orbits its host with a period of the order of several days. The observational data are used to derive radii, masses, orbital parameters and metallicities of the components of the ten binary systems. A subsequent section is dedicated to comparisons of the measured radii with predictions by stellar evolution models. 
 
The ten very-low-mass stars significantly increase the number of known close binary objects with mass and radius measurements in the fully convective regime, $M_\star \lesssim 0.35$ M$_\odot$. We obtain a mean precision in the mass and radius of 4.2\% and 7.5\% respectively. Several systems have lower uncertainties in radius, depending on the quality of the transit photometry. The brightness of the primary star prevents a detection of the radial velocity signature of the low-mass companion, a knowledge of the primary star mass is then required to derive the companion mass and radius. Here, the primary star mass was inferred from stellar evolution models for solar-type stars, using spectroscopic measurements of the metallicity and effective temperature, and the density that was obtained from the model of the eclipse and radial velocity data. 

This requirement for invoking stellar models is a drawback of single-line eclipsing binaries and limits the obtainable precision in the derived radius. We note however that the bright host star permits a measurement of the metallicity of low-mass stars in binary systems that is otherwise difficult to obtain, due to the intrinsically low brightness of low-mass stars. Surface gravities of the low-mass companions can be derived directly from the radial velocity and eclipse data, without invoking stellar models.

%######################################################
%#### FIRST PART - observations, analysis, results ####
%######################################################

% the second part treats comparisons of the results with stellar evolution models

\section{Observations}
The eclipses of the binary stars were detected by the \textit{Wide Angle Search for Planets} \mbox{\citep[WASP, \href{http://wasp-planets.net}{wasp-planets.net};][]{Pollacco_2006}}, a robotic all-sky survey for transiting extra-solar planets. Very-low-mass stars are comparable in size to large Jovian planets and are frequently indistinguishable from large planets in photometric surveys for transiting extra-solar planets. The objects we present here were identified as non-planetary false positives by measuring the host star radial velocity variation, from spectra that were obtained with the fibre-fed CORALIE échelle-spectrograph \citep{CORALIE_HARPS_QUELOZ}. The stars were flagged for further observation, to study the properties of very-low-mass stars in the context of the EBLM project \citep{EBLM_1}. The coordinates, magnitudes and \textit{Gaia} parallax measurements of the target stars are listed in Table \ref{Targets-coordinates}. Intermittent observations by WASP were conducted over several years, and candidate transit events were identified by the automated \texttt{Hunter} algorithm \citep{Collier_Cameron_2007}.

\noindent Spectra of the host stars were obtained with the high-resolution CORALIE spectrograph \citep{CORALIE_HARPS_QUELOZ, EBLM_4} mounted on the \textit{Euler} telescope. In one case, data obtained by the HARPS spectrograph were used. The radial velocity measurements used in this paper were initially reported in \citet{EBLM_4}. The 120-cm \textit{Leonhard Euler} \citep{Lendl_2013} and 60-cm TRAPPIST \citep{Gillon_2011, Jehin_2011} telescopes were used to obtain follow-up photometry of transit events with better precision than the WASP discovery light curves. The data reduction for the two instruments is described in \citet{Lendl_2012} and \citet{Delrez_2014} for \textit{Euler} and TRAPPIST, respectively. In one case (EBLM J1115-36), the SPECULOOS facility \citep{Burdanov_2017,Delrez_2018} was used to obtain a follow-up eclipse observation. Some follow-up observations captured only incomplete eclipses, if the transit coincided with the beginning or end of a night, or was interrupted by adverse weather. To be included in the analysis, we required that in such cases the eclipse shape and depth were well constrained by a fit in the Bayesian framework described in the following section. For two systems, WASP discovery light curves were included in the fit to improve the constraint on the eclipse duration. The eclipse light curves and radial velocity measurements for each system are provided in the appendix. 

\begin{table}
	\renewcommand{\arraystretch}{1.75}
	\centering
	\caption{Coordinates, magnitudes, and parallax measurements of the host stars, determined by the \textit{Gaia} mission \citep{gaia_2016,Gaia_2018b}.}
	\label{Targets-coordinates}
    \begin{adjustbox}{max width=0.49\textwidth}
	\begin{tabular}[width=1\textwidth]{ll|lrrr}
		\hline \hline 
		& EBLM & Coordinates [RA-Dec] & \textit{G} [mag] & Plx [mas]  \\
		\hline 		
		& J0339+03A & 033909.63+030537.5 & 11.42 & 3.30$^{+0.04}_{-0.04}$ & \\
		& J1038-37A & 103824.51--375018.1 & 13.50 & 1.41$^{+0.02}_{-0.02}$ & \\
		& J0555-57A & 055532.69--571726.0& 10.00\tablefootmark{*} & 4.73$^{+0.03}_{-0.03}$ & \\
		& J0954-23A & 095452.89--231955.7 & 10.52 & 3.60$^{+0.10}_{-0.10}$ & \\
		& J0543-56A & 054351.45--570948.5 & 11.69 & 3.05$^{+0.02}_{-0.02}$ & \\
		& J1013+01A & 101350.84+015928.1 & 11.21 & 5.86$^{+0.06}_{-0.06}$ & \\
		& J1115-36A & 111559.67--362733.9 & 12.30 & 1.45$^{+0.03}_{-0.03}$ & \\
		& J1403-32A & 140340.20--323327.3 & 11.96 & 3.89$^{+0.08}_{-0.08}$ & \\
		& J1431-11A & 143152.15--111840.4 & 12.55 & 2.42$^{+0.08}_{-0.08}$ & \\
		& J2017+02A & 201735.83+021551.0 & 11.31 & 3.70$^{+0.05}_{-0.05}$ & 
	\end{tabular}
    \end{adjustbox}
	\tablefoot{\tablefoottext{*}{EBLM J0555-57AB is a visual binary system, where J0555-57A is itself a spectroscopic binary.}}
\end{table}

\section{Spectroscopic analysis of the host-stars}

The spectroscopic analysis of the host stars was performed using a wavelet decomposition of the stellar spectrum, to distinguish spectral features from noise. The spectra enable measurements of the effective stellar temperature, $T_{\rm eff}$, surface gravity, $\log g$, metallicity, [Fe/H], and the sky-projected rotation velocity, $v \sin i_1$, where $i_1$ denotes the inclination of the stellar spin axis. The wavelet analysis is implemented in the \texttt{waveletspec} python package, developed for the analysis of CORALIE spectra of F- and G-type WASP target stars by \cite{Gill2017}. The method is briefly described here: the spectra were co-added and re-sampled between 450 nm and 650 nm, with $2^{17}$ sample points. The wavelet coefficients were then calculated and fitted with coefficients from model spectra in a Bayesian framework, by sampling the posterior distributions of $T_{\rm eff}$, [Fe/H], $v\sin\,i_1$ and $\log g$ of the model spectra. A correction of +0.18 dex was applied to the measurement of the metallicity, to account for a systematic offset of the method with respect to a benchmark sample identified by \citet{Gill2017}. The authors also note a correlation of $\log g$ with $T_{\rm eff}$ that was corrected for using Eq. 9 provided in \citet{Gill2017}. The wavelet method for CORALIE spectra can determine $T_{\rm eff}$ up to a precision of $85$\,K, [Fe/H] to a precision of 0.06\,dex and $v \sin i_1$ to a precision of 1.35 km\,s$^{-1}$ for stars with $v\sin i_1 > 5$ km\,s$^{-1}$. Conservative uncertainties of 124 K, $0.14$ dex and 1.35 km\,s$^{-1}$ were adopted here. Details of the wavelet method and comparisons with other tools commonly used for the spectroscopic analysis of sun-like stars are provided in \citet{Gill2017}. 

\section{Bayesian model of the data}

The eclipse observations and radial velocities were modelled in a fully Bayesian framework. Best-fit parameters were inferred by sampling the posterior probability distribution for the parameters of a generative model for the eclipse and radial velocity measurements. The \texttt{ellc} binary star model \citep{ellc} was used to compute radial velocities and model fluxes for transits.

\subsection{Radial velocities}

The primary star radial velocity determines the orbital period $P$, eccentricity $e$, and longitude of periastron $\omega$, of the companion star. The semi-amplitude of the primary star radial velocity is given by,
\begin{equation}
K = \frac{2\pi a_1 \sin i}{P(1-e^2)^{1/2}},
\end{equation}
where $a_1$ denotes the primary star semi-major axis and $i$ denotes the inclination of the companion star orbit. The semi-amplitude can be related to the component masses by Kepler's third law, 
\begin{equation}
K^3 =  \frac{2\pi G\ m_2^3\sin^3i}{P(1-e^2)^{3/2}(m_1+m_2)^2} =  \frac{2\pi G}{P(1-e^2)^{3/2}} \cdot f_m.
\label{mass-function}
\end{equation}
\noindent The mass function, $f_m$, is defined by,
\begin{equation}
f_m = \frac{m_2^3 \sin^3 i }{(m_1 + m_2)^2},
\end{equation}
for component masses $m_1$ and $m_2$. Expression \ref{mass-function} implies that if the orbital inclination $i$ and primary mass $m_1$ are known, the companion mass $m_2$ can be solved for numerically. The orbital inclination can be determined from the eclipse geometry.

\subsection{Eclipses}

The \texttt{ellc} routine \citep{ellc} computes the orbital configuration at every timestep, integrates over the visible area of the stellar disc, and returns normalized fluxes. 
The transiting very-low-mass companion is assumed to have a negligible luminosity. The maximum fractional reduction of the brightness of the host star, $D$, is then $D = R_2^2/R_1^2$, where $R_1$ and $R_2$ denote the component radii. The eclipse signal is parametrized using the eclipse depth, $D$, the eclipse duration, $W$, and impact parameter $b$. For circular orbits these parameters are related geometrically to the component radii and the orbital inclination $i$, by,
\begin{equation}
b = \frac{a}{R_1}\cos i 
\label{impact parameter}
\end{equation}
\begin{equation}
W = \frac{P}{\pi}\arcsin\left(\frac{R_1}{a}\left\lbrace\frac{\left(1+\frac{R_2}{R_1}\right)^2 - \left[\frac{a}{R_1}\cos i\right]^2}{\sin^2i}\right\rbrace^{\frac{1}{2}}\right),
\label{width}
\end{equation}
where $a$ denotes the semi-major axis of the system \citep{Winn_2011}. Expressions \ref{impact parameter} and \ref{width} depend on the stellar radii only via $r_1 := R_1 / a$ and $r_2 := R_2 / a$. The parameters $r_1$, $r_2$, and $i$ are arguments of the \texttt{ellc} routine that is used to calculate the eclipse model. For eccentric orbits, first-order correction factors for expressions \ref{impact parameter} and \ref{width} were adopted from \citet{Winn_2011}, to account for the dependence of the orbital velocity and orbital separation on the orbital phase. 

The effect of limb-darkening on the shape of the eclipse signal was modelled  using a two-parameter quadratic limb-darkening law implemented in the \texttt{ellc} routine. The limb-darkening parameters were interpolated from the table by \citet{Claret_2004}, using the spectroscopic measurements of temperature, surface gravity and metallicity. 

\subsection{Sampling of the posterior probability distribution}

\noindent A Markov chain Monte Carlo method (MCMC) was used, to sample the joint posterior probability distribution of the transit and radial velocity parameters,
\begin{equation}
p\left(\overline{\mu}\ |\ \overline{f}_{\rm obs}, \overline{v}_{\rm obs}\right) \propto \mathcal{L}\left({\overline{f}_{\rm obs}, \overline{v}_{\rm obs}} |\ \overline{\mu}\right)\ \pi(\overline{\mu}).\label{jointposterior}
\end{equation}
The model parameters are denoted by \hbox{{$\overline{\mu} = \{D,\ W,\ b,\ t_0,\ P,\ K,\ e,\ \omega,\ \gamma,\ a,\ b\}$}}, where $t_0$ is the mid-transit time, and $a$ and $b$ are limb-darkening parameters. The radial velocity measurements and transit photometry are denoted by $\overline{v}_{\rm obs}$ and $\overline{f}_{\rm obs}$ respectively. The prior probability distribution, $\pi(\overline{\mu})$, is uninformative in all parameters, unless stated otherwise. In the case of the system EBLM J0555-57, a prior on the transit-depth was used to account for blending of the transit photometry by a third star; details are provided in \citet{Boetticher_2017}. For all objects, uniform priors were used to constrain the eccentricity to $e \in [0,1)$ and impact parameter to $b \in [0,1]$. Gaussian priors were used to constrain the limb-darkening parameters to the values interpolated from \citet{Claret_2004}.

To account for a possible underestimate of the level of uncorrelated noise in the radial velocities, for instance due to stellar jitter on timescales much shorter than the observation timescale, a parameter was included in the fit that adjusts the uncertainties to $\sigma_{\rm{rv}} \mapsto \sigma_{\rm{rv}} + j$. 
Similarly, a scaling parameter was introduced for the uncertainties of the normalised flux measurements, $\sigma_{\rm{f}} \mapsto \sigma_{\rm{f}} \times s$. The scaling parameters, $s$, $j$, were free parameters in the sampling, and were therefore marginalised. The logarithmic likelihood function used in the MCMC sampling is then given by,
\begin{align}
\ln \mathcal{L}\left(\overline{f}_{\rm obs}, \overline{v}_{\rm obs}\ |\ \overline{\mu}\right) \label{likelihoodfunc}
= &-\frac{1}{2} \sum_i \left[\chi_{f,i}^2 + \ln(\sigma_{f,i}^2)\right] \\ \nonumber &- \frac{1}{2}\sum_i \left[\chi_{v,i}^2 + \ln(\sigma_{v,i}^2)\right],
\end{align}
where the uncertainties $\sigma_{f,i}$ and $\sigma_{v,i}$ are now dependent on the scaling parameters, $s$ and $j$. In (\ref{likelihoodfunc}),\ $\chi_{f,i}^2 = (f_{\rm{obs}, i} - f_i)^2 / \sigma_{f,i}^2$, where the flux measurement is denoted by $f_{\rm obs, i}$, and the computed model flux is denoted by $f_i$. The notation is analogous for the radial velocity measurements in $\chi_{v,i}^2$.  

An initial fit of the data with free scaling parameters for the uncertainties was performed to determine $s$ and $j$. The uncertainties were then re-scaled using $s$ and $j$, and were fixed at their re-scaled values. The residuals of the fit of the light curves were then analysed for time-correlated noise: the residuals were binned for a range of bin widths around the bin width corresponding to the timescale of the eclipse ingress and egress. The presence of time-correlated noise was identified by a deviation of the residuals from the $1/\sqrt{n}$-scaling that is expected for uncorrelated noise, for $n$ residuals per bin. In the absence of a model for correlated noise, the effect of such noise was accounted for by increasing the uncertainties by a factor equal to the quotient of the maximum RMS deviation of the binned residuals, and the RMS deviation of the non-binned residuals \citep{Pont_2005}. The method is described in detail in \citet{Winn_2008,Gillon_2012}, and compared with other common treatments of correlated noise in \citet{Cubillos_2016}. The factor of increase of the photometric uncertainties, after accounting for uncorrelated and correlated noise, was found to lie between one and two for all objects. A final fit of the data was performed using fixed uncertainties, subject to the adjustments to account for uncorrelated and correlated noise. 

\noindent The eccentricity and longitude of periastron were re-parametrised as $f_s = \sqrt{e}\sin \omega$ and $f_c = \sqrt{e}\cos \omega$, to improve the sampling efficiency at very low eccentricities, when $\omega$ is poorly constrained, while maintaining a uniform prior on the value of $e$ \citep[see e.g.][]{Ford_2005,triaud_w23}.
Baselines for transit light curves and radial velocities were fitted with a least-squares algorithm. For transit observations, baselines up to third order in time, up to second order in the flux centroid position on the telescope detector, and up to second order in the flux centroid FWHM were tested. A subtraction of the background flux was performed for some transits, when the transit ingress or egress coincided with dusk or dawn. To evaluate the relative likelihood of baselines of varying complexity, the Bayesian information criterion  \citep[BIC;][]{Schwarz_1978} was used. An improvement of the BIC of 6 was demanded to justify a model of increased complexity \citep{Kass_1995}. 
We fitted a circularised orbit, $e = 0$, to all radial velocities, and compared this fit to an eccentric model using the Bayesian information criterion. Where a circularised fit was imposed as a result of this comparison, this is indicated in Tables \ref{Spectral-results} and \ref{Spectral-results-2}. Two systems, EBLM J0543-56 and EBLM J1038-37, required temporal baselines of second or third order in the radial velocity fit to account for a drift of the radial velocities that may be induced by a third orbiting component. Fitting a superposition of two Keplerian orbits for the two objects, to model the effect of a third component, was not supported by the Bayesian information criterion. Further radial velocity measurements distributed over longer time periods are required to rule out or confirm such a third component in the two systems. The fit baselines used for each system are listed in Tables \ref{Spectral-results} and \ref{Spectral-results-2}. Where an eclipse observation was incomplete and provided a poor constraint on the eclipse duration, the WASP discovery light curve was also used in the fit. The depth of the eclipse determined from the WASP photometry can be unreliable due to the automatic detrending algorithm used by WASP; to account for this, a free parameter was introduced to adjust the eclipse depth of the WASP photometry to that of the TRAPPIST or \textit{Euler} observation.

The joint posterior distribution of the parameters (\ref{jointposterior}) was sampled using an affine invariant stretch-move MCMC algorithm \citep{Goodman_2010}, implemented in the \texttt{emcee} routine \citep{emcee}. An ensemble of 100 'walkers' was sequentially evolved in parameter space for 10 000 steps. Steps are proposed for a walker by extrapolating a line between the present position of the walker, and the position of a randomly chosen other walker. In this way, the movement of a walker is informed by the knowledge of the posterior distribution held by other walkers. Details of the implementation and of the stretch-move algorithm are provided in \citet{emcee} and \citet{Goodman_2010}, respectively. The initial 5000 steps of each walker were discarded, to exclude a burn-in phase required for the ensemble of walkers to converge. Convergence of the walkers was checked by inspecting the chains. 
The fraction of accepted steps of the walkers was also monitored. Marginalised posterior distributions were obtained for each parameter by computing histograms, and were used to compute samples for the component radii, masses, densities and surface gravities. The modal values of the posterior distributions are reported with uncertainties at the 68\% confidence level in tables \ref{Spectral-results} and \ref{Spectral-results-2}.

\subsection{Companion surface gravity}

The surface gravity of the low-mass companion stars, $g_2$, can be determined directly from the eclipse and radial velocity parameters \citep{Southworth_2007}, and is given by,
\begin{equation}
g_2 = \frac{Gm_2}{R_2^2} = 2\pi \frac{K(1-e^2)^{1/2}}{P \sin i\ r_2^2}.
\label{surface-gravity}
\end{equation} 
The semi-amplitude $K$, period $P$, and eccentricity $e$, are determined by the radial velocity fit. The scaled companion radius $r_2 = R_2 / a$ and the orbital inclination $i$, are determined by the transit observation. In the mass--radius plane, the surface gravity $g_2 = G m_2 / R_2^2$ then constrains a star to an iso--gravity line, independent of the primary star mass, that was inferred from stellar models for solar-type stars.

\subsection{Inference of primary star mass and age}

The mass and age of the primary star were inferred by interpolating the GARSTEC \citep{Weiss_2007} stellar evolution models for solar-type stars, using the spectroscopic measurements of $T_{\rm eff}$ and [Fe/H], and the primary star density $\rho_1$. The open-source \texttt{bagemass} routine \citep{bagemass} implements the interpolation of the GARSTEC models in a Bayesian framework. The primary star density was determined iteratively: the density was initialised at the solar density, $\rho_1 = \rho_\odot$, and the GARSTEC models were used to determine an estimate for the primary mass, $m_1$. The expression for the mass-function (\ref{mass-function}),
\begin{equation}
\frac{m_2^3 \sin^3 i}{(m_1+m_2)^2} = \frac{(1-e^2)^{3/2}PK^3}{2\pi G}, 
\end{equation}
was then solved numerically for the companion mass, $m_2$, and the mass ratio, $q = m_2 / m_1$, was used to determine the semi-major axis, $a = a_1(1+1/q)$. The primary radius, $R_1 = a r_1$, was used to update the primary density, $\rho_1 = m_1 / (4\pi R_1^3 / 3)$, and the updated density was used for a new interpolation of the GARSTEC models, to refine the estimate of $m_1$. The computation was repeated iteratively until the radius converged to within the radius uncertainties. The GARSTEC models employed in the \texttt{bagemass} iteration use a default mixing-length-parameter $\alpha_{\mathrm{MLT}} = 1.78$, determined by solar calibration \citep{bagemass}. We discuss the effect of the choice of the mixing length parameter on the derived stellar properties in the following section. 

\section{Results}

The physical properties and orbit parameters for each system are reported in Tables \ref{Spectral-results} \& \ref{Spectral-results-2}. Modal values were determined from the marginalised posterior distribution for each parameter by computing histograms. The 68$\%$-level confidence interval is provided as a measure of the uncertainty.  

%%%%%%%%%%%%%% First five objects %%%%%%%%%%%%%%%

\begin{table*}[h!]
\renewcommand{\arraystretch}{1.4}
\begin{center}
\caption{Spectroscopic measurements, MCMC parameters and derived parameters for five host and companion stars. Continued in Table \ref{Spectral-results-2}. Uncertainties correspond to the 68$\%$ confidence level, where these are provided in brackets, they refer to the last significant figures. The instruments used for the observations are indicated by TR - TRAPPIST, Eu - Euler, SP - SPECULOOS, W - WASP, COR - CORALIE spectrograph, HARPS - HARPS spectrograph. Dates refer to BJD$_{\rm UTC}$ $-$ 2 450 000.}
\label{Spectral-results}
\begin{tabular*}{\textwidth}{ll|rrrrrr}
\hline \hline 
&Parameter&J0555-57&J0954-23&J1431-11&J2017+02&J0543-56&\\
\hline 	
\multicolumn{6}{l}{\textit{Spectroscopic measurements of primary star}}\\
&$T_{\rm eff}$ [K] &6368$^{+124}_{-124}$&6406$^{+124}_{-124}$&6161$^{+124}_{-124}$&6161$^{+124}_{-124}$&6223$^{+124}_{-124}$&\\
&[Fe/H] [dex]  &$-0.04$$^{+0.14}_{-0.14}$&$-0.01$$^{+0.14}_{-0.14}$&0.15$^{+0.14}_{-0.14}$&$-0.07^{+0.14}_{-0.14}$&0.23$^{+0.14}_{-0.14}$&\\
&$\log g_1$ [dex] &4.10$^{+0.21}_{-0.21}$&4.15$^{+0.21}_{-0.21}$&4.25$^{+0.21}_{-0.21}$&4.27$^{+0.21}_{-0.21}$&4.38$^{+0.21}_{-0.21}$&\\
&$v \sin i_1$ [km\,s$^{-1}$] &7.30$^{+1.35}_{-1.35}$&7.83$^{+1.35}_{-1.35}$&10.40$^{+1.35}_{-1.35}$&76.70$^{+1.35}_{-1.35}$&$< 5$&\\
\multicolumn{6}{l}{\textit{Parameters from \texttt{bagemass} iteration}}\\
&System age [Gyr] &1.6$^{+1.2}_{-1.2}$&2.0$^{+1.0}_{-1.0}$&1.6$^{+1.3}_{-1.3}$&3.9$^{+1.7}_{-1.7}$&1.1$^{+1.1}_{-1.1}$&\\
& $M_{1}$ [$M_{\odot}$]&1.180$^{+(82)}_{-(79)}$&1.166$^{+(80)}_{-(82)}$&1.200$^{+(56)}_{-(55)}$&1.105$^{+(74)}_{-(72)}$&1.276$^{+(72)}_{-(70)}$\\

\multicolumn{6}{l}{\textit{Free parameters in MCMC sampling}}\\
&$P$ [d]&7.757675$^{+(19)}_{-(18)}$&7.574661$^{+(15)}_{-(16)}$&4.450156$^{+(06)}_{-(06)}$&0.82178768$^{+(47)}_{-(48)}$&4.4638602$^{+(21)}_{-(20)}$&\\
&$t_0$ [d]& 6712.6449$^{+(12)}_{-(12)}$& 7872.5198$^{+(12)}_{-(13)}$&7871.62566$^{+(59)}_{-(62)}$&7468.4936$^{+(11)}_{-(11)}$&7716.77724$^{+(62)}_{-(64)}$&\\
&$D$&0.00475$^{+(18)}_{-(17)}$*&0.00624$^{+(52)}_{-(52)}$&0.01785$^{+(52)}_{-(50)}$&0.0164$^{+(13)}_{-(13)}$&0.02366$^{+(69)}_{-(69)}$&\\
&$W$ [d]&0.1386$^{+(35)}_{-(25)}$&0.1257$^{+(74)}_{-(68)}$&0.1420$^{+(21)}_{-(18)}$&0.0893$^{+(22)}_{-(18)}$&0.1594$^{+(22)}_{-(21)}$&\\
&$b$&0.41$^{+(21)}_{-(24)}$&0.827$^{+(44)}_{-(72)}$&0.00$^{+(16)}_{-(00)}$&0.28$^{+(18)}_{-(17)}$&0.00$^{+(15)}_{-(00)}$&\\
&$K$ [km\,s$^{-1}$]&7.739$^{+(29)}_{-(30)}$&8.6903$^{+(74)}_{-(76)}$&13.004$^{+(36)}_{-(37)}$&26.68$^{+(18)}_{-(72)}$&16.650$^{+(12)}_{-(12)}$\\
&$f_s$&0.2428$^{+(89)}_{-(86)}$&$-0.1943^{+(28)}_{-(28)}$&0 (fixed)&0 (fixed)&0 (fixed)&\\
&$f_c$& $-0.1767^{+(53)}_{-(53)}$&$-0.0629^{+(39)}_{-(38)}$&0 (fixed)&0 (fixed)&0 (fixed)&\\

\multicolumn{6}{l}{\textit{Derived parameters}}\\
&$M_1$ [M$_\odot$]&1.180$^{+(82)}_{-(79)}$&1.166$^{+(80)}_{-(82)}$&1.200$^{+(56)}_{-(55)}$&1.105$^{+(74)}_{-(72)}$&1.276$^{+(72)}_{-(70)}$\\
&$R_1$ [R$_\odot$]&1.00$^{+(14)}_{-(07)}$&1.23$^{+(17)}_{-(17)}$&1.114$^{+(43)}_{-(28)}$&1.196$^{+(80)}_{-(50)}$&1.255$^{+(54)}_{-(36)}$&\\
&$\rho_1$ [$\rho_\odot$]&1.23$^{+(20)}_{-(33)}$&0.53$^{+(33)}_{-(19)}$&0.884$^{+(48)}_{-(89)}$&0.654$^{+0.0650}_{-0.0198}$&0.649$^{+0.040}_{-0.072}$\\
&$M_2$ [M$_\odot$]&0.0839$^{+(38)}_{-(38)}$&0.0981$^{+(56)}_{-(57)}$&0.1211$^{+(36)}_{-(37)}$&0.1357$^{+(63)}_{-(64)}$&0.1641$^{+(57)}_{-(59)}$&\\
&$R_2$ [$R_\odot $]&0.0844$^{+(131)}_{-(60)}$&0.101$^{+(17)}_{-(17)}$&0.1487$^{+(70)}_{-(50)}$&0.153$^{+(13)}_{-(10)}$&0.1929$^{+(100)}_{-(70)}$&\\
&$\log g_2$ [cgs]&5.51$^{+0.06}_{-0.12}$&5.41$^{+0.15}_{-0.13}$&5.18$^{+0.03}_{-0.04}$&5.20$^{+0.05}_{-0.07}$&5.09$^{+0.03}_{-0.04}$&\\
&$f_m$ [M$_\odot$]&0.0003685$^{+(43)}_{-(44)}$&0.0005138$^{+(13)}_{-(14)}$&0.0010140$^{+(83)}_{-(88)}$&0.001616$^{+0.000031}_{-0.000127}$&0.0021350$^{+(44)}_{-(44)}$&\\
&$e$&0.0895$^{+(35)}_{-(36)}$&0.04186$^{+(94)}_{-(92)}$&0 (fixed)&0 (fixed)&0 (fixed)&\\
&$a$ [au]&0.0828$^{+0.0018}_{-0.0019}$&0.0820$^{+0.0018}_{-0.0019}$&0.05815$^{+0.00086}_{-0.00088}$&0.01850$^{+0.00038}_{-0.00040}$&0.0600$^{+0.0011}_{-0.0011}$&\\
&$\omega$ [deg]&$-54.0^{+1.6}_{-1.7}$&72.1$^{+1.2}_{-1.2}$&180 (fixed)&180 (fixed)&180 (fixed)\\
&$i$ [deg]&89.3$^{+0.9}_{-1.1}$&86.84$^{+(66)}_{-(61)}$&89.95$^{+(56)}_{-(84)}$&86.6$^{+3.1}_{-3.6}$&89.99$^{+(75)}_{-(92)}$ \\
&$P_{\rm rot,\ primary}$ [d]**& 6.92$^{+1.61}_{-1.61}$&7.95$^{+1.76}_{-1.76}$&5.40$^{+0.72}_{-0.72}$&0.79$^{+0.06}_{-0.06}$&-&\\
\multicolumn{6}{l}{\textit{Instruments and fit properties}}\\
&Instruments& TR, Eu, COR & TR, COR & TR, COR &TR, W, COR & TR, HARPS, COR & \\
& Radial velocity baseline & - & - & - & - & quadratic time \\
& Flux baseline & $-$background$^{\dagger}$ & - & - & $-$background& - &
\end{tabular*}
\vspace{0.25cm}
\tablefoot{The detection of the sky-projected rotation velocity is reliable for $v\sin i_1 > 5\ \rm{km\,s^{-1}}$. The high precision of the orbital period of EBLM J2017+02Ab and EBLM J1115-36Ab is due to the use of the WASP discovery light curves in the fit, together with radial velocities and transits observed with TRAPPIST. *$D$ denotes the observed eclipse depth of EBLM J0555-57A. A blend of the eclipse by a third component is accounted for in the derivation of the stellar radii, described in \citet{Boetticher_2017}. **Assuming a negligible inclination of the primary star spin axis. $^{\dagger}$Subtraction of the background flux.}
\end{center}
\end{table*}

%%%%%%%%%%%%%% Second five objects %%%%%%%%%%%%

\begin{table*}[h!]
\renewcommand{\arraystretch}{1.4}
\begin{center}
\caption{Continuation of Table \ref{Spectral-results}. Spectroscopic measurements, MCMC parameters and derived parameters for the second five host and companion stars. Uncertainties correspond to the 68$\%$ confidence level, where these are provided in brackets, they refer to the last significant figures. The instruments used for the observations are indicated by TR - TRAPPIST, Eu - Euler, SP - SPECULOOS, COR - CORALIE spectrograph. Dates refer to BJD$_{\rm UTC}$ $-$ 2 450 000.}
\label{Spectral-results-2}
\begin{tabular*}{\textwidth}{ll|rrrrrr}
\hline \hline 
&Parameter&J1038-37&J1013+01&J1115-36&J0339+03&J1403-32&\\
\hline 	
\multicolumn{6}{l}{\textit{Spectroscopic measurements of primary star}}\\
&$T_{\rm eff}$ [K] &5885$^{+124}_{-124}$&5570$^{+124}_{-124}$&6605$^{+124}_{-124}$&6132$^{+124}_{-124}$&5826$^{+124}_{-124}$&\\
&[Fe/H] [dex]  &0.31$^{+0.14}_{-0.14}$&0.29$^{+0.14}_{-0.14}$&0.30$^{+0.14}_{-0.14}$&$-0.25^{+0.14}_{-0.14}$&0.19$^{+0.14}_{-0.14}$&\\
&$\log g_1$ [dex] &4.50$^{+0.21}_{-0.21}$&4.68$^{+0.21}_{-0.21}$&4.06$^{+0.21}_{-0.21}$&4.00$^{+0.21}_{-0.21}$&4.51$^{+0.21}_{-0.21}$&\\
&$v \sin i_1$ [km\,s$^{-1}$] &$< 5$&16.09$^{+1.35}_{-1.35}$&11.53$^{+1.35}_{-1.35}$&17.00$^{+1.35}_{-1.35}$&$< 5$&\\

\multicolumn{6}{l}{\textit{Parameters from \texttt{bagemass} iteration}}\\
&System age [Gyr] &4.8$^{+2.1}_{-2.1}$&5.4$^{+2.6}_{-2.6}$&1.9$^{+0.5}_{-0.5}$&5.8$^{+2.0}_{-2.0}$&1.8$^{+1.5}_{-1.5}$&\\
&$M_1$ [M$_\odot$]&1.176$^{+(72)}_{-(70)}$&1.036$^{+(70)}_{-(72)}$&1.369$^{+(72)}_{-(72)}$&1.036$^{+(74)}_{-(76)}$&1.083$^{+(50)}_{-(51)}$&\\

\multicolumn{6}{l}{\textit{Free parameters in MCMC sampling}}\\
&$P$ [d]&5.021614$^{+(16)}_{-(10)}$&2.8922726$^{+(24)}_{-(24)}$&10.5426599$^{+(14)}_{-(16)}$&3.580673$^{+(11)}_{-(11)}$&11.908745$^{+(76)}_{-(76)}$&\\
&$t_0$ [d]& 6289.6986$^{+(22)}_{-(41)}$&5741.7889$^{+(17)}_{-(17)}$&7644.96592$^{+(79)}_{-(77)}$&6129.5891$^{+(11)}_{-(11)}$&6834.4958$^{+(16)}_{-(15)}$& \\
&$D$&0.03279$^{+(83)}_{-(79)}$&0.04312$^{+(45)}_{-(44)}$&0.01504$^{+(34)}_{-(31)}$&0.0295$^{+(10)}_{-(10)}$&0.0847$^{+(17)}_{-(17)}$&\\
&$W$ [d]&0.1364$^{+(85)}_{-(53)}$&0.1263$^{+(13)}_{-(13)}$&0.2459$^{+(41)}_{-(42)}$&0.1468$^{+(37)}_{-(34)}$&0.1760$^{+(31)}_{-(30)}$&\\
&$b$&0.561$^{+(42)}_{-(64)}$&0.003$^{+(92)}_{-(03)}$&0.23$^{+(11)}_{-(12)}$&0.36$^{+(09)}_{-(13)}$&0.008$^{+(54)}_{-(08)}$&\\
&$K$ [km\,s$^{-1}$]&17.645$^{+(29)}_{-(30)}$&23.194$^{+(81)}_{-(83)}$&13.011$^{+(76)}_{-(76)}$&24.849$^{+(50)}_{-(59)}$&20.938$^{+(11)}_{-(11)}$&\\
&$f_s$& 0 (fixed) &0 (fixed)&$-0.101^{+(29)}_{-(27)}$&0 (fixed)&$-0.2971^{+(13)}_{-(13)}$&\\
&$f_c$& 0 (fixed) &0 (fixed)&0.203$^{+(12)}_{-(12)}$&0 (fixed)&0.1416$^{+(25)}_{-(23)}$&\\

\multicolumn{6}{l}{\textit{Derived parameters}}\\
&$M_1$ [M$_\odot$]&1.176$^{+(72)}_{-(70)}$&1.036$^{+(70)}_{-(72)}$&1.369$^{+(72)}_{-(72)}$&1.036$^{+(74)}_{-(76)}$&1.083$^{+(50)}_{-(51)}$&\\
&$R_1$ [R$_\odot$]&1.132$^{+(52)}_{-(48)}$&1.036$^{+(27)}_{-(26)}$&1.579$^{+(48)}_{-(41)}$&1.210$^{+(55)}_{-(52)}$&0.969$^{+(21)}_{-(20)}$&\\
&$\rho_1$ [$\rho_\odot$]&0.793$^{+(93)}_{-(96)}$&0.940$^{+(35)}_{-(42)}$&0.351$^{+(20)}_{-(27)}$&0.576$^{+(72)}_{-(65)}$&1.197$^{+(59)}_{-(56)}$&\\
&$M_2$ [M$_\odot$]&0.1735$^{+(66)}_{-(67)}$&0.1773$^{+(75)}_{-(77)}$&0.1789$^{+(61)}_{-(59)}$&0.2061$^{+(93)}_{-(95)}$&0.2755$^{+(77)}_{-(79)}$&\\
&$R_2$ [R$_\odot$]&0.205$^{+(11)}_{-(10)}$&0.2150$^{+(60)}_{-(60)}$&0.1929$^{+(80)}_{-(60)}$&0.207$^{+(12)}_{-(11)}$&0.2824$^{+(80)}_{-(80)}$&\\
&$\log g_2$ [cgs]&5.04$^{+0.04}_{-0.04}$&5.02$^{+0.01}_{-0.02}$&5.12$^{+0.02}_{-0.03}$&5.12$^{+0.05}_{-0.05}$&4.98$^{+0.02}_{-0.02}$&\\
&$f_m$ [M$_\odot$]&0.002859$^{+(14)}_{-(14)}$&0.003739$^{+(39)}_{-(40)}$&0.002400$^{+(42)}_{-(42)}$&0.005697$^{+(35)}_{-(40)}$&0.011130$^{+(17)}_{-(17)}$&\\
&$e$& 0 (fixed)&0 (fixed)& 0.0522$^{+(38)}_{-(37)}$& 0 (fixed)&0.10820$^{+(67)}_{-(67)}$&\\
&$a$ [au]&0.0634$^{+0.0012}_{-0.0013}$&0.04238$^{+0.00087}_{-0.00092}$&0.0492$^{+0.0011}_{-0.0012}$&0.0492$^{+0.0011}_{-0.0012}$&0.1132$^{+0.0016}_{-0.0016}$&\\
&$\omega$ [deg]& 0 (fixed)&0 (fixed)&$-26.0^{+(7.6)}_{-(7.1)}$&0 (fixed)&$-64.62^{+(44)}_{-(43)}$\\
&$i$ [deg]&87.38$^{+(30)}_{-(24)}$&89.98$^{+(49)}_{-(61)}$&89.19$^{+(47)}_{-(46)}$&87.67$^{+(90)}_{-(65)}$&89.961$^{+0.090}_{-0.136}$&\\
&$P_{\rm rot,\ primary}$ [d]*& -&3.26$^{+0.35}_{-0.35}$&6.93$^{+0.84}_{-0.84}$&3.60$^{+0.33}_{-0.33}$&-&\\
\multicolumn{6}{l}{\textit{Instruments and fit properties}}\\
&Instruments& Eu, COR & TR, COR & TR, SP, W, COR & TR, COR & Eu, COR \\
&Radial velocity baseline& cubic time & -  & - & - & - & \\
&Flux baseline& linear time & - & - & linear time & lin. centroid-$y$** &
\end{tabular*}
\tablefoot{*Assuming a negligible inclination of the primary star spin axis. **Linear baseline in the $y$-coordinate of the centroid position on the detector.}
\end{center}
\end{table*}

The mass--radius posterior distributions for each object are shown in Figure \ref{MR_posteriors}, visualised using kernel density estimates, with 68$\%$-level confidence contours shown in blue. The mass and radius posterior distributions are correlated, due to the dependence of the derived radius on the primary mass, $R_2 = r_2 \cdot a(m_1, m_2(m_1))$. The mass-log$g_2$ posterior distributions are shown in Figure \ref{Mlogg_posteriors}. The measurements are uncorrelated, since the surface gravity can be derived without assuming a primary star mass.

\begin{figure*}
    \begin{minipage}[t][][t]{\columnwidth}
        \centering
	    \includegraphics[width=\linewidth]{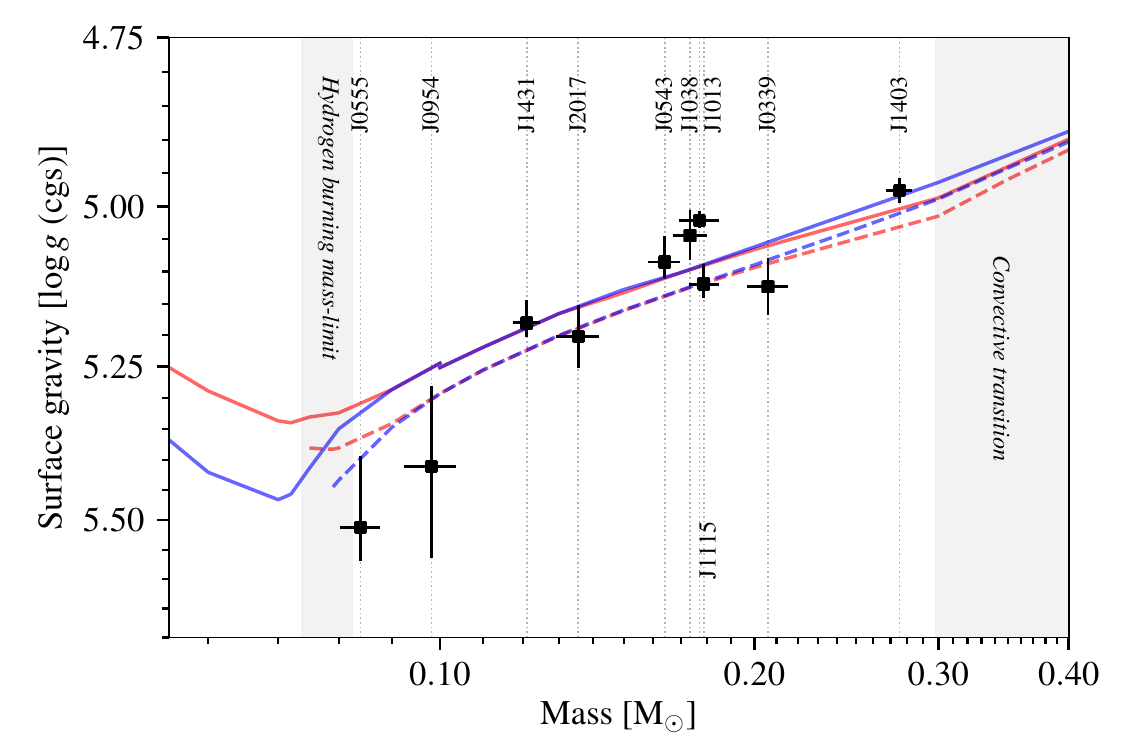}
	    \caption{Measurements of the surface gravities of the ten very-low-mass stars, plotted against the stellar mass. The surface gravity is derived independently of the primary star mass, measurements of the surface gravity and mass are therefore uncorrelated. The Exeter/Lyon \citep{Baraffe_2015, Baraffe_98, Baraffe_2003} 1 Gy (red) and 5 Gy (blue) isochrones are shown, for solar metallicity, [Fe/H] = 0.0 dex (solid line), and sub-solar metallicity, [Fe/H] = $-0.5$ dex (dashed line).}
	    \label{fig:MASS-LOGG}
    \end{minipage}
    \hfill
    \begin{minipage}[t][][t]{\columnwidth}
    	\centering 
	    \includegraphics[width=\linewidth]{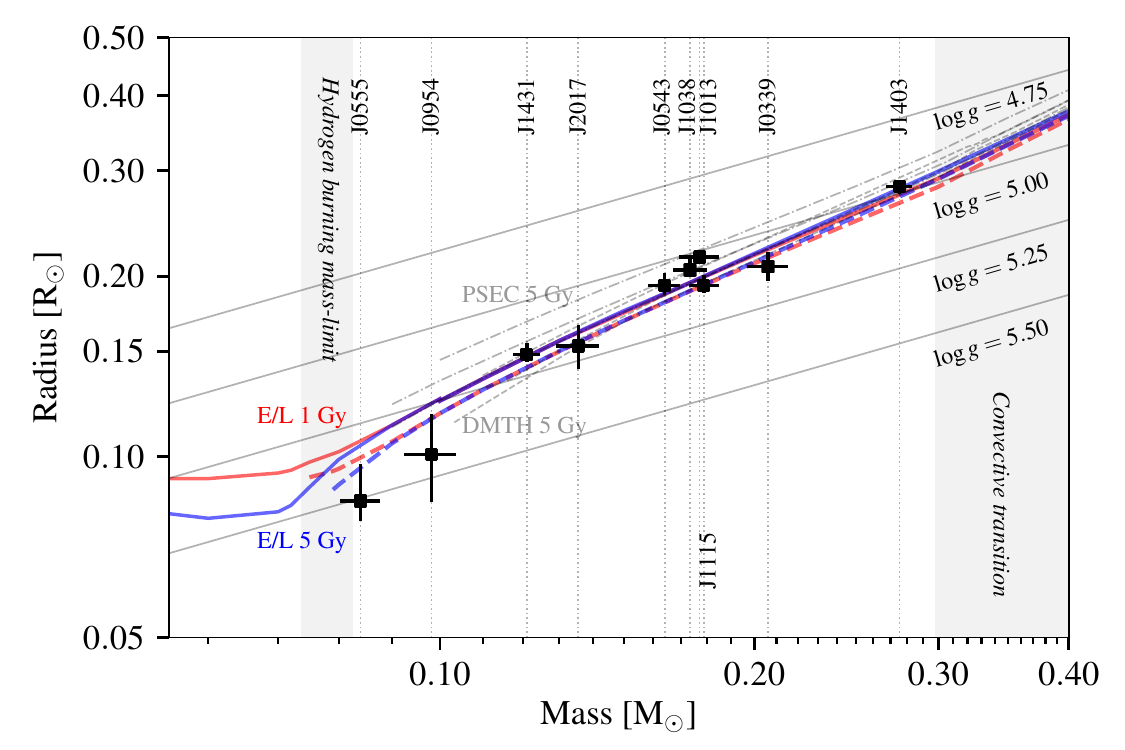}
	    \caption{Masses and radii of the ten very-low-mass stars. The Exeter/Lyon  \citep[E/L;][]{Baraffe_2015, Baraffe_98, Baraffe_2003} 1 Gy (red) and 5 Gy (blue) isochrones are shown, for solar metallicity, [Fe/H] = 0.0 dex (solid lines), and sub-solar metallicity, [Fe/H] = $-0.5$ dex (dashed lines). The Dartmouth  \citep[DMTH;][]{Dotter_2008} (dashed, grey) and PARSEC  \citep[PSEC;][]{Marigo_2017} isochrones (dot-dashed, grey) are also shown; for sub-solar (smallest radii), solar, and super-solar ([Fe/H] = +0.5 dex) metallicity (largest radii).}
	    \label{fig:MASS-RADIUS}
    \end{minipage}
\end{figure*}

The mass--radius diagram in Figure \ref{fig:MASS-RADIUS} shows the modal positions of the ten objects in the very-low-mass, fully convective regime, $\sim$0.078 M$_{\odot} \lesssim $ $M$  $\lesssim 0.3 $ M$_\odot$. The Exeter/Lyon (E/L) \citep{Baraffe_98, Baraffe_2015} 1 and 5 Gy isochrones are shown for solar metallicity, [Fe/H] = 0.0 dex, and for sub-solar metallicity (dashed lines), [Fe/H] $= -0.5$ dex. The hydrogen-burning minimum mass lies approximately at 0.07 M$_\odot$ \citep{Kumar_1963} and is indicated in grey. We note that the hydrogen-burning minimum mass is a function of the metallicity; lower metallicities imply a higher hydrogen-burning minimum mass \citep{Chabrier_2000}. The transition region from stars that are fully convective to stars that have a radiative core is indicated in grey at $0.3-0.4$ M$_\odot$. Figure \ref{fig:MASS-LOGG} shows the mass-surface gravity diagram. The relative uncertainties in the surface gravity are not significantly lower than uncertainties in the radius, indicating that uncertainties in the stellar radii are dominated by the quality of the photometry, and not by the uncertainty associated with the primary mass estimate from stellar models. In the following sections we make comparisons with stellar models in the mass-radius plane. 

\subsection{Notable objects and spin-orbit synchronisation}

The stars EBLM J0555-57Ab and EBLM J0954-23Ab are comparable in size to Jovian planets. EBLM J0555-57Ab was previously characterised in \citet{Boetticher_2017}; the results presented here reflect updated spectroscopic measurements, that correct for a fault in the data reduction pipeline of the CORALIE spectrograph. Radial velocities were unaffected by this correction and the derived masses and radii are consistent with previous results. 

The radius of EBLM J1013+01Ab is the most precisely determined in the sample of ten stars, and exceeds the radius predicted by the Exeter/Lyon stellar evolution models. The joint posterior distribution of mass and radius is shown in Figure \ref{MR_posteriors}. The interpolation of the GARSTEC models implies an age of $5.4 \pm 2.6$ Gyr
for the primary star, suggesting that a very young age of the system cannot be invoked to explain the large radius. Isochrones from the PARSEC \citep{Marigo_2017} and Dartmouth stellar evolution models \citep{Dotter_2008} are also plotted in Figure \ref{fig:MASS-RADIUS}, up to [Fe/H] = +0.5 dex, and indicate better consistency. The PARSEC models incorporate empirically motivated adjustments to the outer boundary conditions of the stellar model, to match the observational properties of very-low-mass stars \citep{Chen_2014}. The spectroscopic measurement of the metallicity of J1013+01A indicates [Fe/H] $= 0.11 \pm 0.14$ dex, consistent with a near-solar metallicity. 

The two systems EBLM J2017+02 and EBLM J1013+01 have the shortest orbital periods in the sample, with $P = 0.82$ and $P = 2.89$ days, respectively. EBLM J2017+02A has a spectroscopically measured projected rotation velocity $v\sin i = 76.71 \pm 1.35$ km\,s$^{-1}$. Using the measurement of the stellar radius and assuming a negligible inclination of the stellar spin axis, this velocity corresponds to a rotation period $P_{\rm rot} = 0.79 \pm 0.06$ days, consistent with the companion orbital period. A synchronization of the primary component spin and the companion orbital period is also found for EBLM J1013+01. The spectroscopic measurement of the projected rotation velocity of the host star is $v\sin i = 16.09 \pm 1.35$ km\,s$^{-1}$. The corresponding rotation period of $3.26 \pm 0.35$ days suggests a near-synchronisation of the stellar spin and companion orbit, assuming a negligible inclination of the stellar spin axis. The rotation periods of the primary stars EBLM J0339+03A, EBLM J1431-11A, EBLM J0555-57A and EBLM J0953-23A are similarly consistent with the orbital periods of their low-mass companions. The derived values for the rotation periods of all objects are provided in Tables \ref{Spectral-results} and \ref{Spectral-results-2}, if a measurement of the rotation velocity was possible.

Tidally induced fast rotation may affect the structure of the primary stars, an effect that is not accounted for in the non-rotational GARSTEC models used to infer the masses of the primary components. For the fastest rotating objects, EBLM J2017+02A and J1013+01A, we investigated the effect of a reduced mixing-length-parameter in the GARSTEC models, to simulate a diminished convective efficiency \citep[][]{Chabrier_2007}. The properties of the components were re-derived with $\alpha_{\mathrm{MLT}} = 1.50$, reduced from the default $\alpha_{\mathrm{MLT}} = 1.78$. For J1013+01Ab, this results in an increased primary mass, $M_1 = 1.11 \pm 0.06$ M$_\odot$, and companion mass and radius $M_2 = 0.1846^{+(63)}_{-(65)}$ M$_\odot$, $R_2 = 0.2196^{+(52)}_{-(50)}$ R$_\odot$, corresponding to an increase of the companion radius by 2.3$\%$. For J2017+02, an increased primary mass, $M_1 = 1.16 \pm 0.07$ M$_\odot$, and companion mass and radius, $M_2 = 0.1393^{+(60)}_{-(63)}$ M$_\odot$, $R_2 = 0.1548^{+(130)}_{-(94)}$ R$_\odot$ are obtained, corresponding to an increase of the companion radius by 1.2$\%$. Neglecting effects of fast stellar rotation may therefore lead to underestimates of the derived radii. To test for such underestimates, the derived primary radii were compared with \textit{Gaia} DR2 radius estimates. The \textit{Gaia} radii are determined from \textit{Gaia} three-band photometry and parallax measurements, using extremely randomized trees; the procedure is described in \citet{Gaia_Apsis}. Typical uncertainties are 10\%, adopted here for all objects, but we note that the \textit{Gaia} astrometric goodness of fit is poor for EBLM J0954-23A, J0555-57A, J0543-56A and J2017+02A; the goodness of fit parameter for these objects $\mathrm{gof_{AL}} > 8$. For good fits, the parameter is normally distributed with $\mathrm{gof_{AL}} \sim \mathcal{N}(0,1)$. For all objects, $\mathrm{gof_{AL}} > 3$, indicating that the \textit{Gaia} radius estimates should be treated with some caution. Comparisons of the primary radii with \textit{Gaia} estimates are shown in Figure \ref{fig:DR2-COMP}, indicating generally good agreement but suggesting possible underestimates of the radii of EBLM J0954-23, J0555-57 and J0339+03. Temperature comparisons are shown in Figure \ref{fig:DR2-COMP2}. Future \textit{Gaia} data releases and a larger sample may enable a reliable comparison. We note that surface gravity measurements of the low-mass components are independent of the primary mass estimate.

% ###############################################################
% #### SECOND PART, COMPARISON WITH STELLAR EVOLUTION MODELS ####
% ###############################################################

\section{Comparisons with stellar evolution models}

The measured radii were compared with theoretical radii predicted by stellar evolution models. The Exeter/Lyon models \citep{Baraffe_2015,Baraffe_2003,Baraffe_98} were used in the comparison because they cover the complete mass range of low-mass stars, down to the hydrogen-burning mass limit. The Dartmouth \citep{Dotter_2008} and PARSEC \citep{Marigo_2017} models do not extend to sufficiently low masses to include EBLM J0954-23Ab and EBLM J0555-57Ab; we used the Dartmouth \citep{Dotter_2008} models as a second benchmark for objects in the mass range of the Dartmouth models. 
\subsection{The mass-radius relation}

Residuals of the radius measurements were calculated with respect to radii predicted by stellar models, given the mass, metallicity and age of the star. The Exeter/Lyon isochrones shown in Figure \ref{fig:MASS-RADIUS} indicate that for the radius uncertainties encountered here the evolution of the stellar radius is negligible between ages of 1 and 5 Gyr, for fully convective low-mass stars of near-solar metallicity. We verified that this holds for ages up to 10 Gyr. We assumed that the primary and secondary component in each binary system are coeval and in the following sections use 5 Gyr isochrones to compare empirical radii with model predictions. For EBLM J0543-57Ab, EBLM J0555-57Ab, EBLM J1403-32Ab and EBLM J1431-11Ab an age $<1$ Gyr cannot be ruled out at the 1-$\sigma$ level.

Residuals with respect to a solar metallicity isochrone, ([Fe/H] = 0.0 dex), are plotted as a function of mass in Fig. \ref{fig:radius_residuals-NOT-ADJUSTED}. Stars with a super-solar metallicity lie above the solar-isochrone, conversely stars with a sub-solar metallicity lie below the solar isochrone. To account for the effect of the metallicity on the stellar radius, we assumed that the metallicities of the low-mass companion stars are identical to the metallicities measured spectroscopically for the primary stars. 
To obtain a predicted radius as a function of metallicity, a linear interpolation was performed between isochrones of sub-solar and solar metallicity. No super-solar metallicity isochrones are available within the Exeter/Lyon models so we extrapolated linearly into the super-solar regime. The model for the stellar radius, as a function of mass and metallicity can then be expressed as, 
\begin{equation}
R_{\mathrm{[Fe/H]}}(m) = R_{0.0}(m) + 2\mathrm{[Fe/H]}(R_{0.0}(m) - R_{-0.5}(m)),
\label{Interpolation}
\end{equation}
where $R_{\mathrm{[Fe/H]}}(m)$ denotes the radius predicted by the stellar model, given the mass $m$, and metallicity $[\mathrm{Fe/H}]$; and $R_{-0.5}(m)$ and $R_{0.0}(m)$ denote the radii predicted by sub-solar ([Fe/H] = $-0.5$ dex) and solar ([Fe/H] = 0.0 dex) metallicity isochrones respectively. The interpolation (Eq. \ref{Interpolation}) was performed for each point in the sample of the mass posterior distribution that was generated by the MCMC sampling. 
To account for the uncertainty of 0.14 dex associated with the metallicity measurement, the metallicity value used in the interpolation for each point in the mass posterior sample was drawn randomly from a normal distribution $\rm{[Fe/H]} \sim \mathcal{N}(\mathrm{[Fe/H]}, \sigma_{\rm [Fe/H]})$. A posterior distribution for the fractional radius residual of a star was then obtained by subtracting the predicted radius from the empirical radius, and dividing by the empirical radius:
\begin{equation}
	\frac{\Delta R}{R} = \frac{R_{\rm obs}-R_{\rm pred}(m_2,\ \rm{[Fe/H]})}{R_{\rm obs}}.
    \label{residuals}
\end{equation} 
The resulting distribution for the fractional radius residual was used to determine the modal residual and 68$\%$-level uncertainties, shown in Fig.~\ref{fig:radius_residuals-adjusted}. Accounting for the stellar metallicity in the calculation of residuals decreases the chi-squared statistic to $\chi^2 = 16.1 \pm 5.7$, from $\chi^2 =  27.8 \pm 7.5$ for residuals with respect to the solar-metallicity isochrone. A mean radius residual of $\Delta R/R = 1.6 \pm 1.2\%$ was determined for all objects. Super-solar metallicity isochrones from the Dartmouth stellar evolution models \citep{Dotter_2008} were used to verify that extrapolating the Exeter/Lyon models into the super-solar regime is acceptable for the metallicities encountered here. The residuals calculated using both models are provided in Table \ref{Radius-residuals}.

\subsection{Effect of the stellar metallicity on radii}

The mass-radius isochrones in Fig. \ref{fig:MASS-RADIUS} show the scaling of the radius with the stellar metallicity. A decreased stellar metallicity yields a lower opacity,  resulting in a higher effective temperature at a given optical depth. Thermodynamic equilibrium implies a higher core temperature, and hydrostatic equilibrium, $R \sim m/T$, requires that the star contracts \citep{Chabrier_2000}. Conversely, metal-enhanced stars have larger radii. We tested for an empirical confirmation of this scaling by calculating radius residuals with respect to the solar-metallicity, $\mathrm{[Fe/H]} = 0.0$ dex, 5-Gy isochrone from the Exeter/Lyon \citep{Baraffe_2015} models.  

\begin{figure*}
    %\begin{minipage}{0.5\textwidth}
    \begin{minipage}[t][][t]{1\columnwidth}
        \centering
	    \includegraphics[width=1\linewidth]{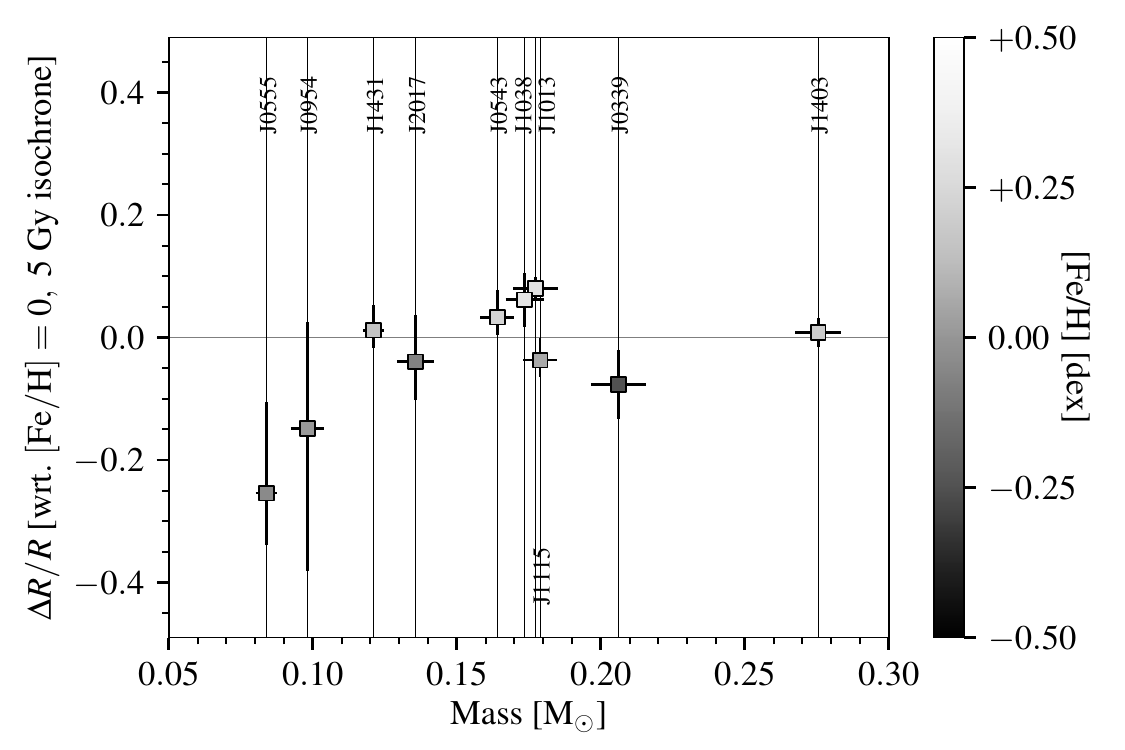}
        \caption{Radius residuals, with respect to the solar [Fe/H] = 0.0 dex, 5--Gy isochrone from \citet{Baraffe_2015}, plotted against the mass. The residuals are not adjusted for the metallicity measured for the primary star. The colour-bar indicates the spectroscopic metallicity measurement of the primary star.}
        \label{fig:radius_residuals-NOT-ADJUSTED}
    \end{minipage}
    \hfill
    \begin{minipage}[t][][t]{1\columnwidth}
        \centering
	    \includegraphics[width=1\linewidth]{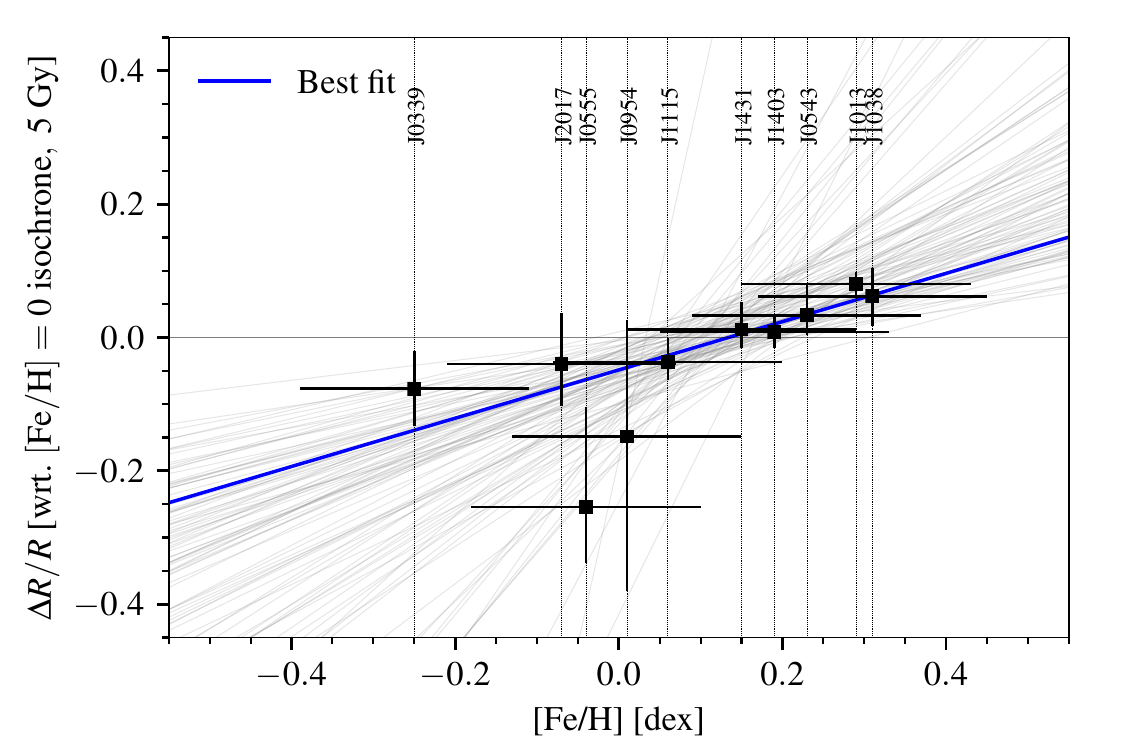}
        \caption{Radius residuals, with respect to the solar, [Fe/H] = 0.0 dex, 5--Gy isochrone \citep{Baraffe_2015}, plotted against the metallicity. The line of best fit is shown in blue, $\Delta R/ R = (0.38^{+0.38}_{-0.18}) \rm{[Fe/H]} - (0.046^{+0.036}_{-0.057})$. Random draws from the posterior distributions of the fit (grey) indicate the uncertainty in the fit parameters.}
        \label{fig:radius_residuals-against_met}
    \end{minipage}
\end{figure*}
\begin{figure*}
     \begin{minipage}[t][][t]{1\columnwidth}
        \centering
        \includegraphics[width=1\linewidth]{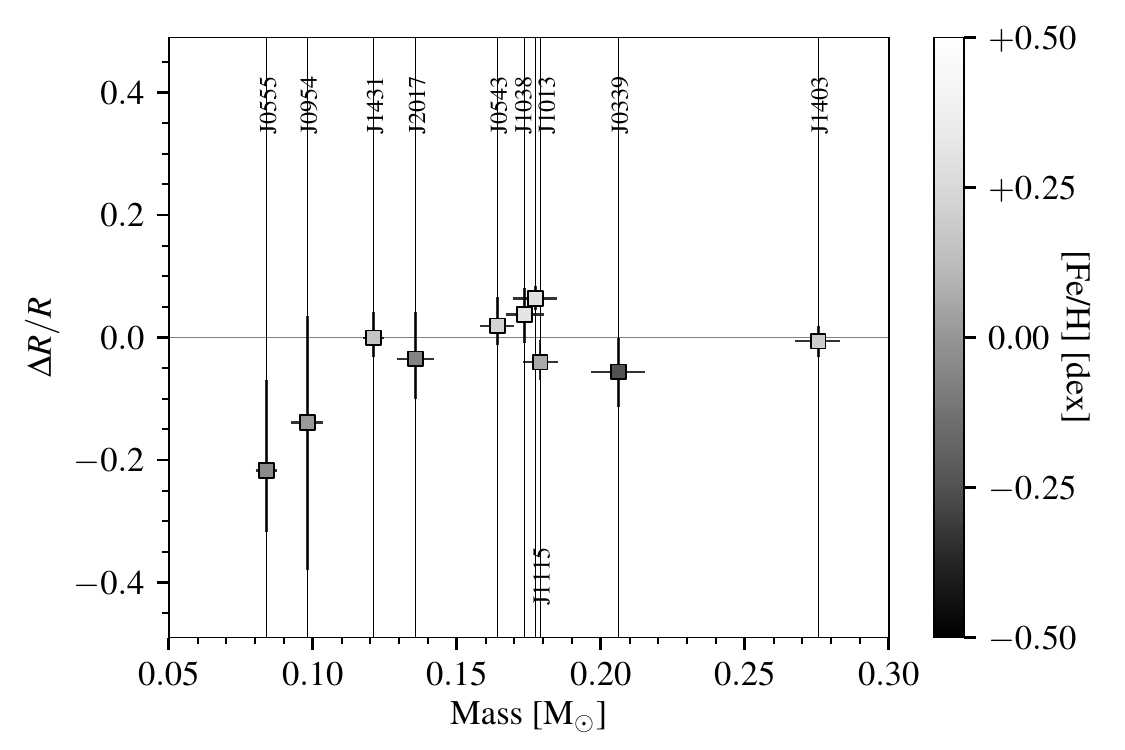} 
	    \caption{Metallicity-adjusted radius residuals, plotted against the mass. Residuals are computed with respect to a predicted radius, determined by using the metallicity measurement for the primary star to linearly interpolate between solar and sub-solar isochrones from \citep{Baraffe_2015,Baraffe_98}, and extrapolating for super-solar metallicities.}
        \label{fig:radius_residuals-adjusted}
    \end{minipage}
    \hfill 
    \begin{minipage}[t][][t]{1\columnwidth}
        \centering
        \includegraphics[width=1\linewidth]{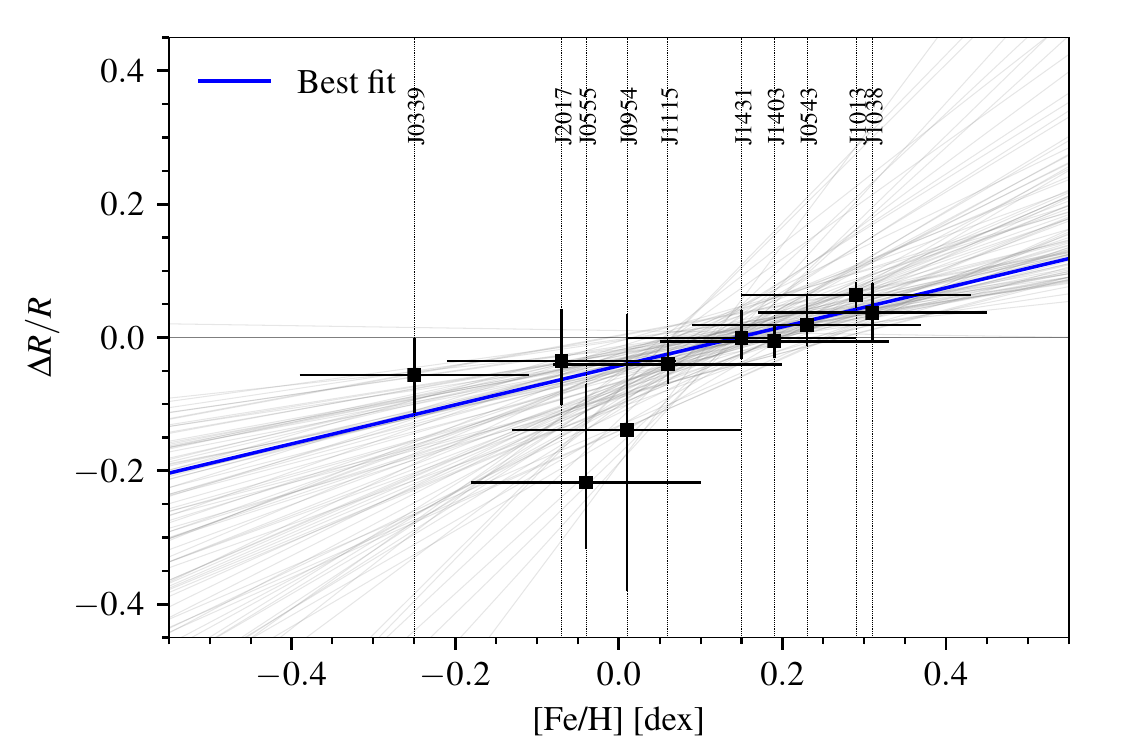}
	    \caption{Metallicity-adjusted radius residuals, plotted against the metallicity. The line of best fit, $\Delta R/ R = (0.29^{+0.33}_{-0.16}) \rm{[Fe/H]} - (0.041^{+0.032}_{-0.051})$ is shown in blue. Random draws from the posterior distributions of the fit (grey) indicate the uncertainty in the fit parameters.}
        \label{fig:radius_residuals-against_met-adjusted}
    \end{minipage}
\end{figure*}

The residuals are shown in Fig.~\ref{fig:radius_residuals-NOT-ADJUSTED} and plotted against the metallicity in Fig. \ref{fig:radius_residuals-against_met}. A test for a correlation between the residuals and the metallicity was performed by sampling the posterior distribution for the parameters of a linear model, $\Delta R/R = m \mathrm{[Fe/H]} +  b$. Residuals in the likelihood function were calculated orthogonal to the linear model to account for uncertainties in both the metallicity measurement and stellar radius. The best fit linear model is
\begin{equation}
\left(\frac{\Delta R}{R}\right)_{\mathrm{[Fe/H]=0.0}} = \left(0.38^{+0.38}_{-0.18}\right) \rm{[Fe/H]} - \left(0.046^{+0.036}_{-0.057}\right).
\end{equation}  
A comparison of the 2-parameter linear fit to a constant fit, $\Delta R/R = c$, using the Bayesian information criterion, determines $\Delta_{\rm BIC} = 15.9 > 6$.  The fit was re-computed if EBLM J1013+01Ab is omitted. The low radius uncertainty of this object and its discrepant position in Fig.~\ref{fig:radius_residuals-against_met-adjusted} suggest that it dominates the correlation. If EBLM J1013+01Ab is ignored, then $\Delta_{\rm BIC} = +4.4 < 6$, over a constant 1-parameter model, confirming that the correlation depends significantly on EBLM J1013+01Ab.  
To examine the effect of accounting for the metallicity in stellar models, the metallicity-corrected residuals are plotted against the metallicity in Fig.~\ref{fig:radius_residuals-against_met-adjusted}. A linear fit determined, 
\begin{equation}
\left(\frac{\Delta R}{R}\right)_{\rm Adj} = \left(0.29^{+0.33}_{-0.16}\right) \rm{[Fe/H]} - \left(0.041^{+0.032}_{-0.051}\right), 
\end{equation}  
where the subscript, '$\mathrm{Adj}$', indicates that the calculation of the residuals accounts for the effect of the stellar metallicity. A comparison of the 2-parameter fit with a 1-parameter constant fit, $\Delta R/R = c$, determines $\Delta_{\rm BIC} = 9.1 > 6$, suggesting that a two-parameter linear model is preferred over the one-parameter, constant model. If EBLM 1013+01Ab is omitted, then $(\Delta R / R)_{\rm Adj} = (0.25^{+0.42}_{-0.19}) \rm{[Fe/H]} - (0.034^{+0.031}_{-0.051})$ and $\Delta_{\rm BIC} = 1.0 < 6$ over a constant model. 
The scaling of the stellar radius with metallicity is in agreement with stellar theory, but we note that the measured correlation is not robust if EBLM 1013+01Ab is omitted. A correlation of radius residuals and metallicity persists after accounting for the metallicity in calculating radius residuals; this correlation likewise relies on EBLM 1013+01Ab. The scaling of the radius with metallicity is consistent with similar findings of \citet{Berger_2006}. 

\subsection{Effect of the orbital period on radii}\label{sec:radius-period}

An inflation of the radii of low-mass stars in binary systems is frequently associated with short orbital periods \citep[eg.][]{Lopez_Morales_2007,Spada_2013}. In short-period binary stars, the fast rotation of tidally locked components can enhance stellar magnetic activity, which may inhibit the convective transport of heat in the stellar interior and cause an inflation of radii \citep{Lopez_Morales_2007, Chabrier_2007}. To examine this hypothesis we tested for a correlation of the radius residuals with the orbital periods of the binary systems.  
Fig. \ref{fig:radius-period} plots the radius residuals against the orbital period. A linear fit over the complete period range determined no significant correlation, $\Delta R/R = (-0.007 \pm 0.013)P - (0.03 \pm 0.09)$. 

Binary systems with orbital periods below $\sim$5 days are typically circularized by tidal effects \citep{Pont_2005}. The components in such circularized binary systems are expected to be tidally locked, such that the rotation period is synchronized with the orbital period for ages exceeding 1 Gy \citep{Zahn_1989, Meibom_2005}. We defined mean radius residuals $\mu_{\rm c}$ and $\mu_{\rm e}$ for circularised ($P \lesssim 5$ days) and eccentric systems ($P \gtrsim 5$ days) respectively, and determined $\mu_{\rm c} = 5.9 \pm 1.5 \%$, and $\mu_{\rm e} = 0.4 \pm 2.1 \%$. We caution that the respective sample sizes of 6 and 4 objects are small, and that EBLM 1013+01Ab dominates the short-period sample due to its low radius uncertainties. Four of the six short-period systems have a super-solar metallicity, making it difficult to disentangle the effect of metallicity and short orbital periods on the stellar radius.

\subsection{Combined analysis with comparable objects}

Comparable low-mass stars from the literature were added to the present sample of ten stars to analyse the radius residuals in an extended sample. We required that all objects are fully convective stars in binary systems, with $0.07$ M$_\odot$ $\lesssim M \lesssim 0.3\ M_\odot$, and that a spectroscopic measurement of the metallicity was available for each system. We identified thirteen objects that match these criteria, listed in Table \ref{Literature-objects}.
Figure \ref{fig:MASS-RADIUS-all} shows the mass-radius diagram for low-mass stars, the thirteen added objects are indicated by blue diamond-shaped markers. Low-mass objects with masses greater than 0.3 M$_\odot$ from \citet{Chen_2014}, and several sub-stellar objects are shown for comparison.

\begin{figure*}
     \begin{minipage}[t][][t]{1\columnwidth}
        \centering
    	\includegraphics[width=1\linewidth]{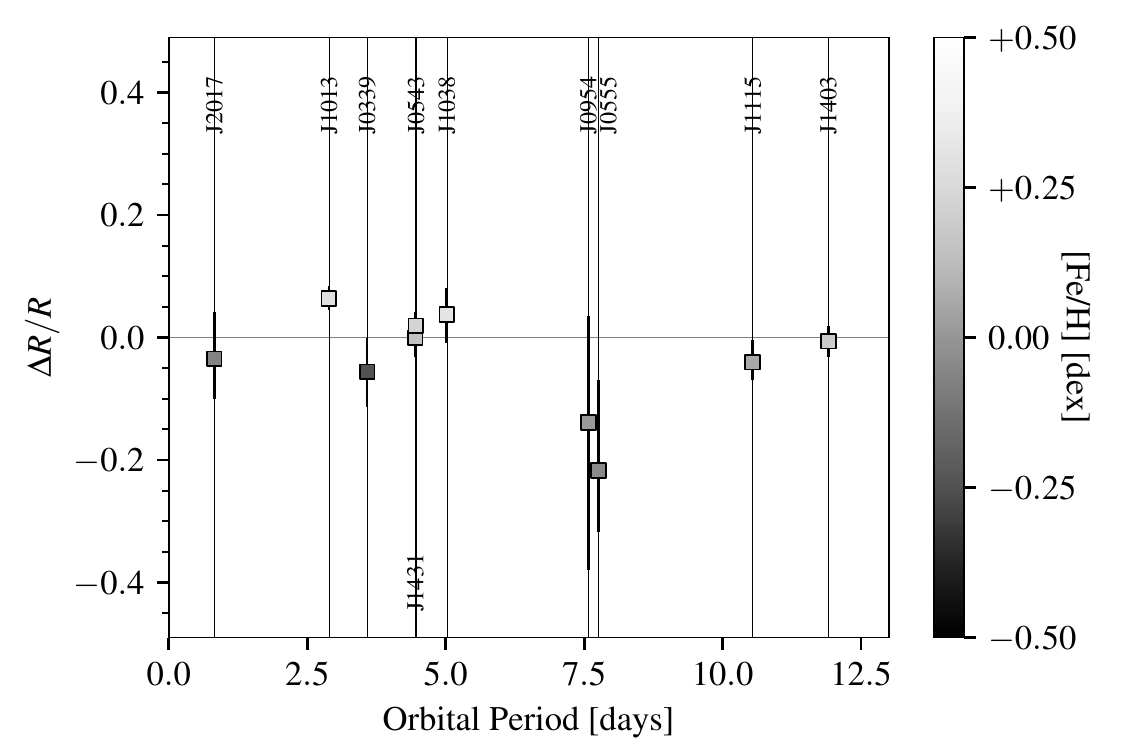}
        \caption{Fractional radius residuals, adjusted for the metallicity, plotted against the orbital period of the low-mass stars. Objects with periods $< 5.1$ days have circularised orbits, $e = 0$. In such systems the rotation periods of the components are expected to be synchronised with the orbital period of the binary.}\label{fig:radius-period}
    \end{minipage}
    \hfill 
    \begin{minipage}[t][][t]{1\columnwidth}
        \centering
        \includegraphics[width=1\linewidth]{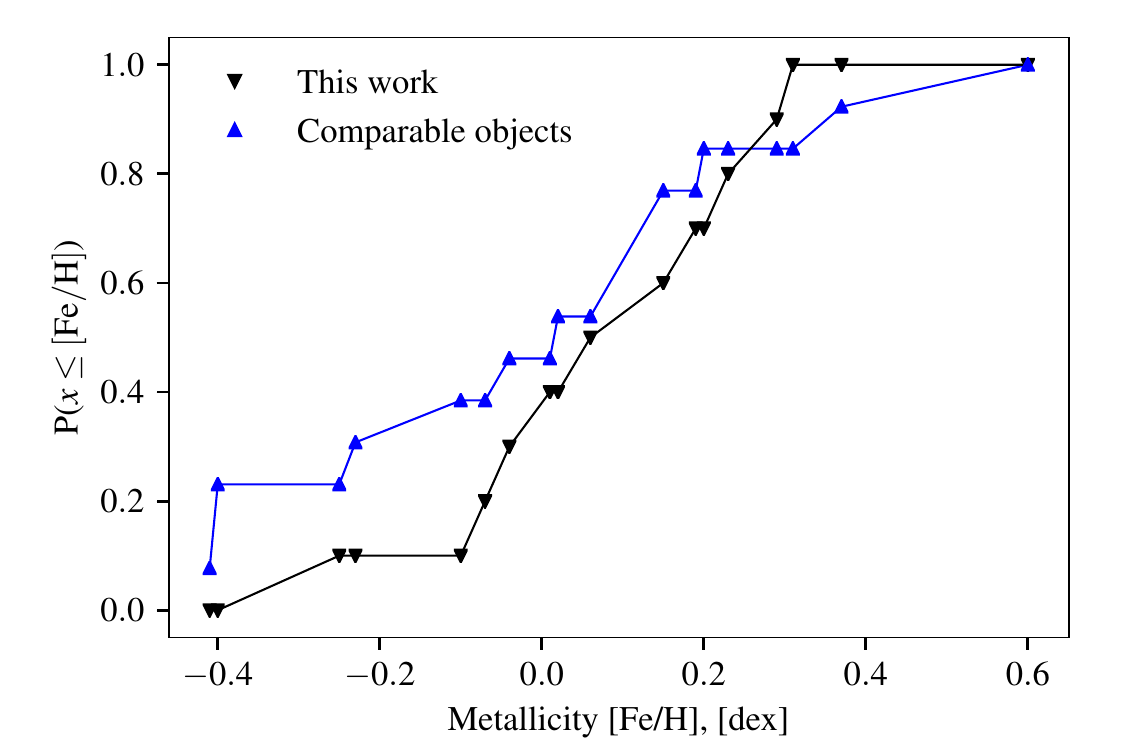}
	    \caption{Empirical distribution functions for the metallicity measurements of the sample of ten stars, and the sample of comparable low-mass stars selected from the literature. The Kolmogorov-Smirnov test does not indicate inconsistency of the metallicity distributions.}
	    \label{fig:metallicity_comparison}
    \end{minipage}
\end{figure*}

\begin{figure*}
	\centering
	\includegraphics[width=\textwidth]{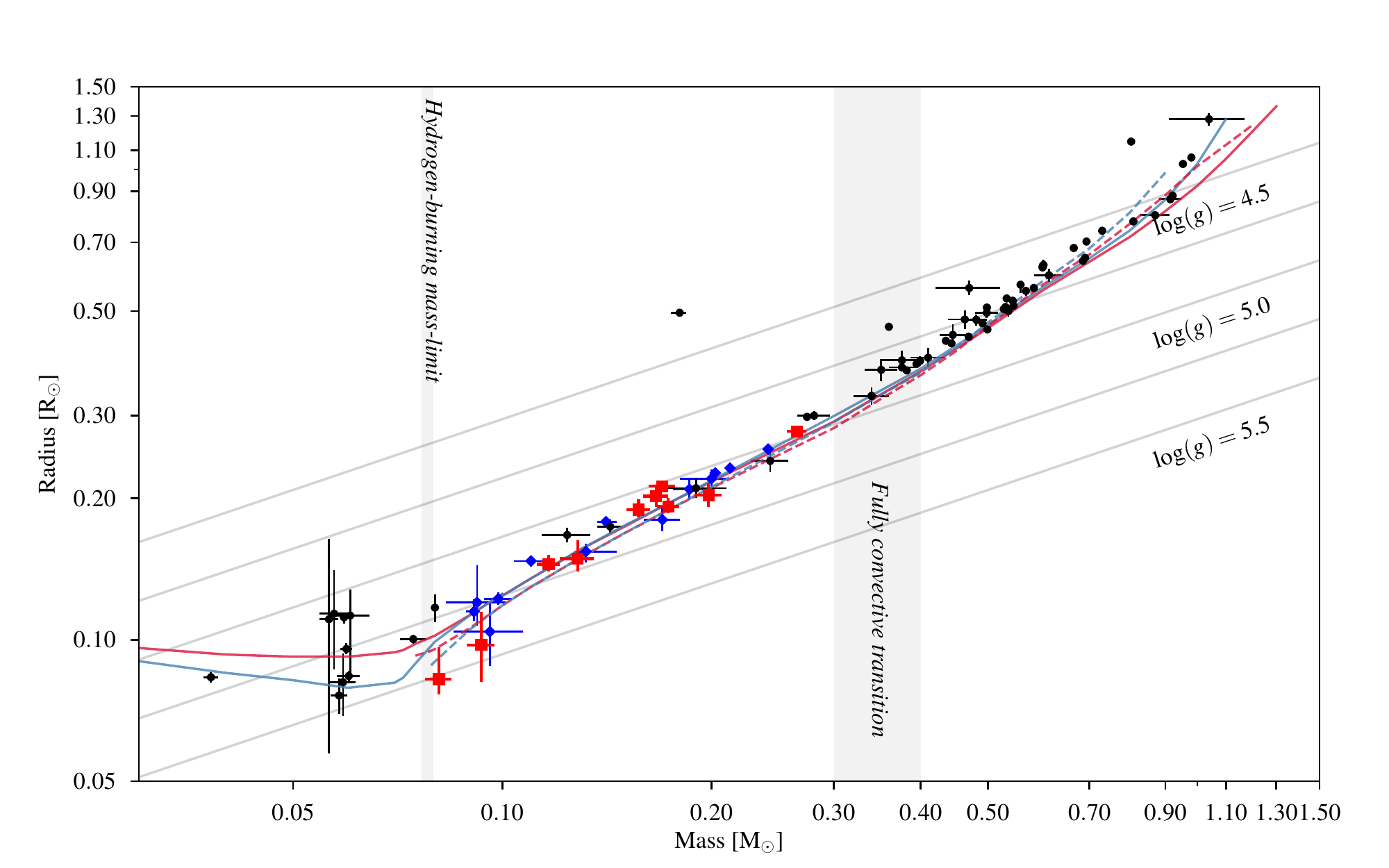}
	\caption{Mass-radius diagram, on logarithmic axes, showing the positions of the ten very-low-mass stars (red, square), binary stars from the literature (black, circular) and comparable literature objects selected to expand the sample (blue, diamonds). The literature objects are detailed in Table \ref{Literature-objects} and \citet{Chen_2014}. Several sub-stellar objects are also shown. The Exeter/Lyon \citep{Baraffe_2015,Baraffe_2003,Baraffe_98} 1 Gy (red) and 5 Gy (blue) isochrones are shown, for solar metallicity [Fe/H] = 0 dex (solid line), and sub-solar metallicity, [Fe/H] = $-0.5$ dex (dashed line). Iso-surface gravity lines are shown for $\log g_2 = 4.5$ dex to $\log g_2 = 5.5$ dex (cgs). Objects are constrained to their surface-gravity line by the $\log g_2$ measurement, derived from the transits and radial velocities without invoking the primary mass.} 
	\label{fig:MASS-RADIUS-all}
\end{figure*}

The metallicities of the current sample of ten stars and the sample of comparable literature objects were compared using the cumulative distribution functions (CDF) of the two metallicity samples. To approximate the CDF, the empirical distribution function of each sample was calculated, on the metallicity domain of the combined sample of 23 stars. The empirical distribution functions are $F_{10}(x) = 1/10 \sum_{i=1}^{10} {s_{x_i < x}}$, and $F_{13}(x) = 1/13 \sum_{i=1}^{13} {s_{x_i < x}}$, for the EBLM and literature sample respectively, where $s_{x_i < x}$ denotes the step function, and the $x_i$ denote the metallicities contained in each sample. The two distributions are shown in Fig. \ref{fig:metallicity_comparison}. A 2-sample Kolmogorov-Smirnov test determined a $p-$value of 0.43 for a test-statistic of 0.28, indicating that the distributions of metallicity of the two samples are not inconsistent.

\begin{figure*}
     \begin{minipage}[t][][t]{1\columnwidth}
        \centering
        \includegraphics[width=1\linewidth]{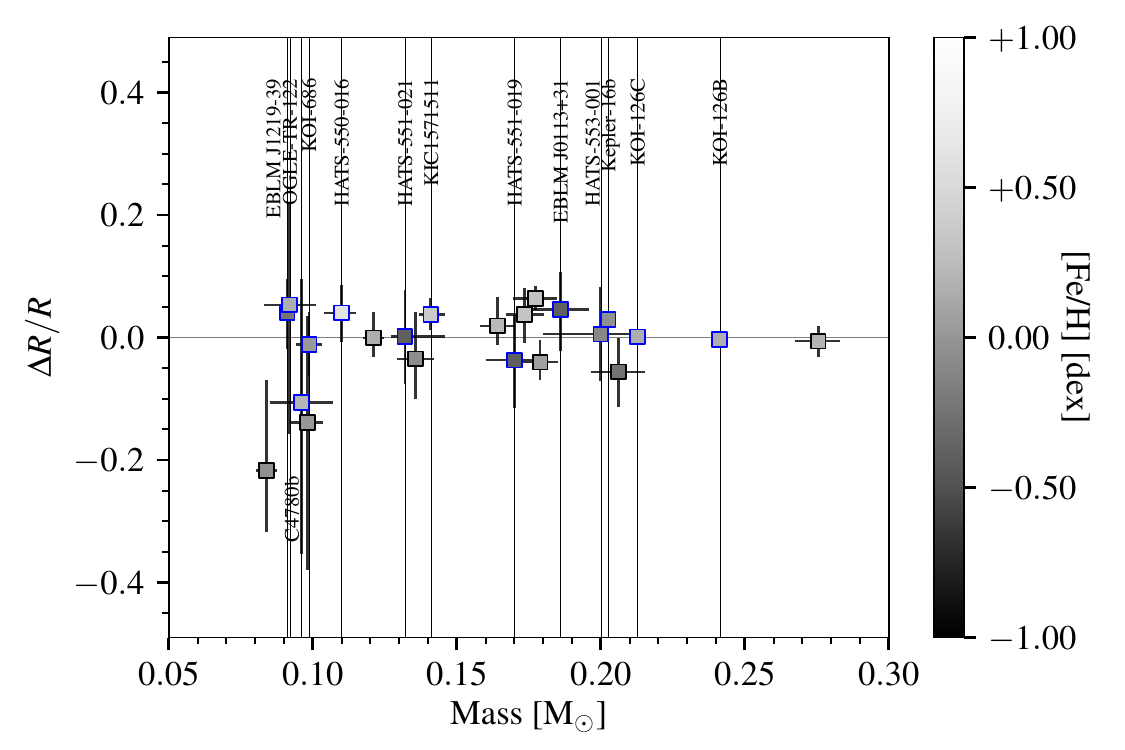}
	    \caption{Radius residuals of the ten low-mass stars and 13 comparable literature objects (squares, blue edge), plotted against the stellar mass.}
	    \label{fig:radius_residuals-adjusted-all}
    \end{minipage}
    \hfill 
    \begin{minipage}[t][][t]{1\columnwidth}
        \centering
    	\includegraphics[width=1\linewidth]{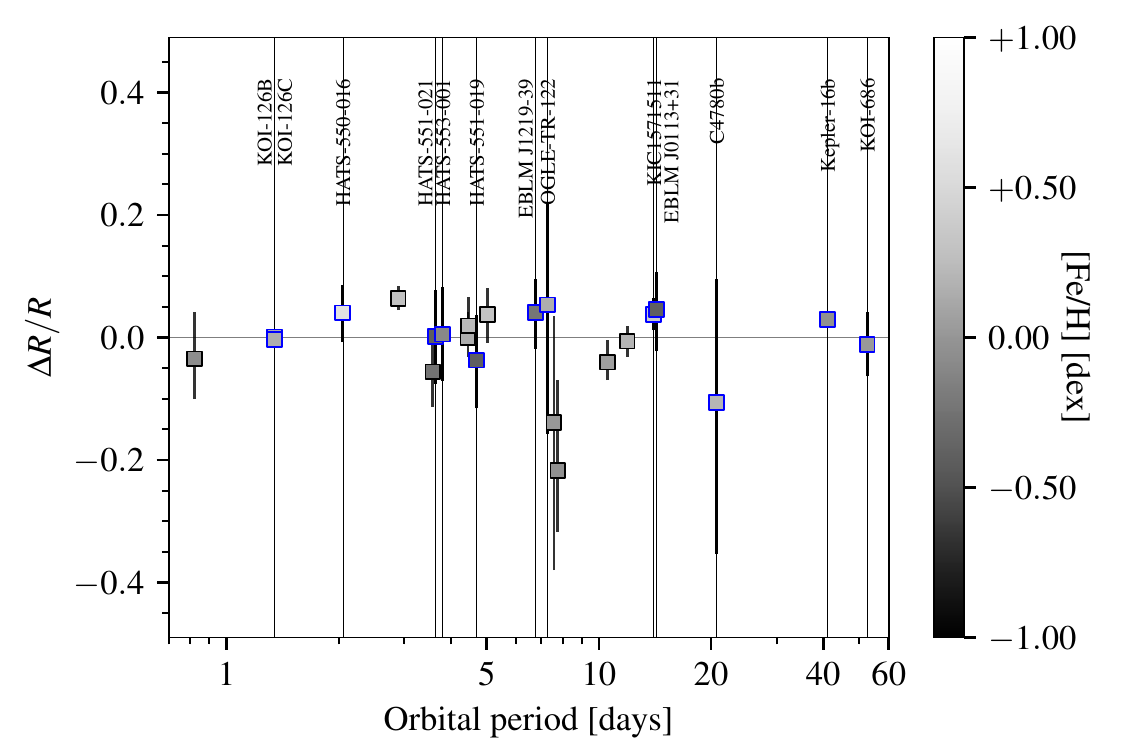}
	    \caption{Radius residuals of the ten low-mass stars and 13 comparable literature objects (squares, blue edge), plotted against the orbital period.}
        \label{fig:radius_residuals-adjusted-all_period}
    \end{minipage}
\end{figure*}

\begin{figure*}[h!]
    \centering
    \begin{minipage}{0.48\textwidth}
        \centering
	    \includegraphics[width=0.99\textwidth]{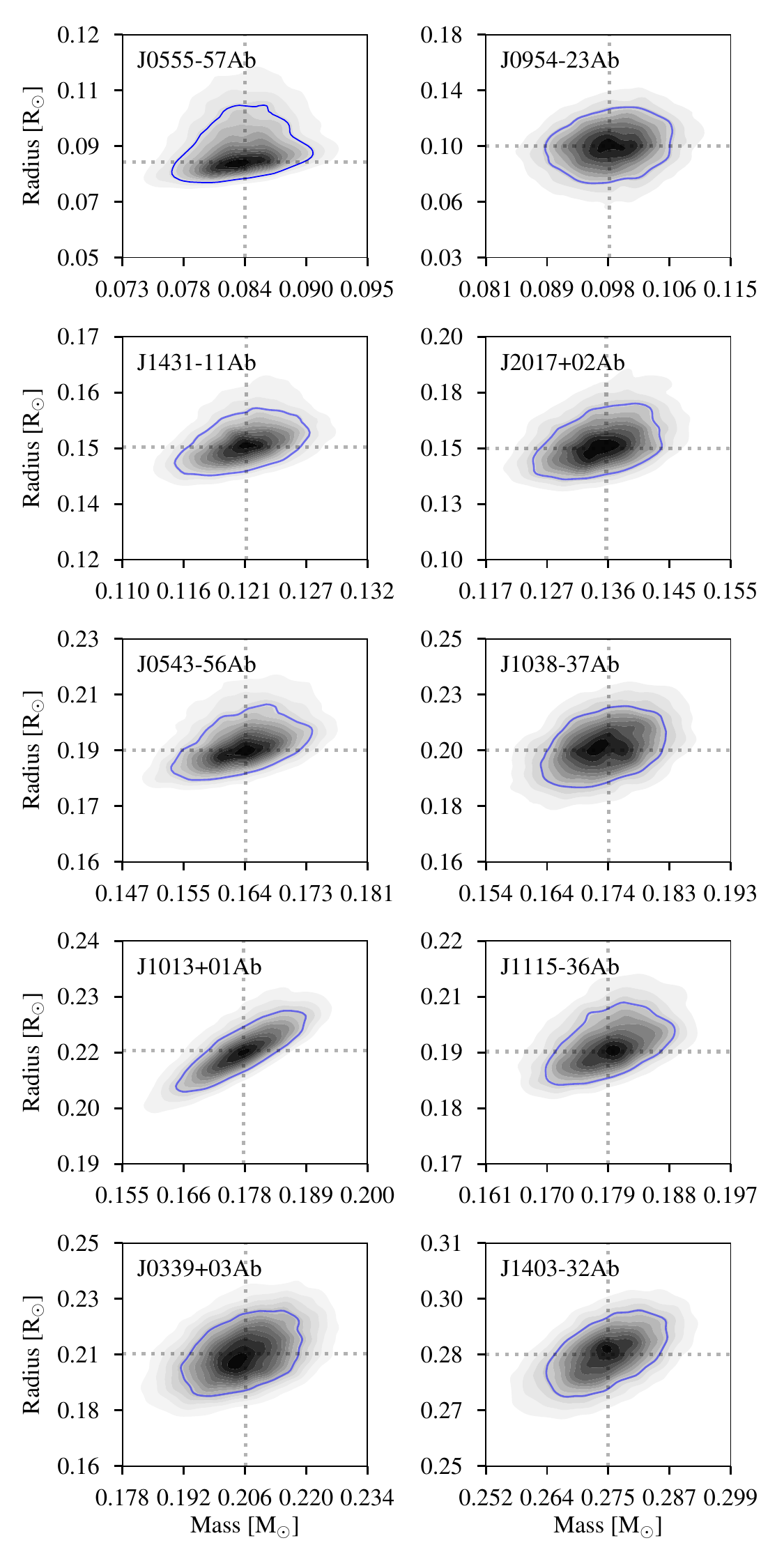}
    	\caption{Mass-radius posterior distributions generated by the Markov chain Monte Carlo sampling. Visualised using kernel-density estimates. The 68$\%$-level contour is shown in blue. Modal values are indicated by dotted lines. The mass and radius posterior distributions are correlated, since the primary star mass estimate is used in the derivation of both parameters.} 
        \label{MR_posteriors}
    \end{minipage}% 
    \hspace{5pt}
    \begin{minipage}{0.48\textwidth}
        \centering
        \includegraphics[width=0.99\textwidth]{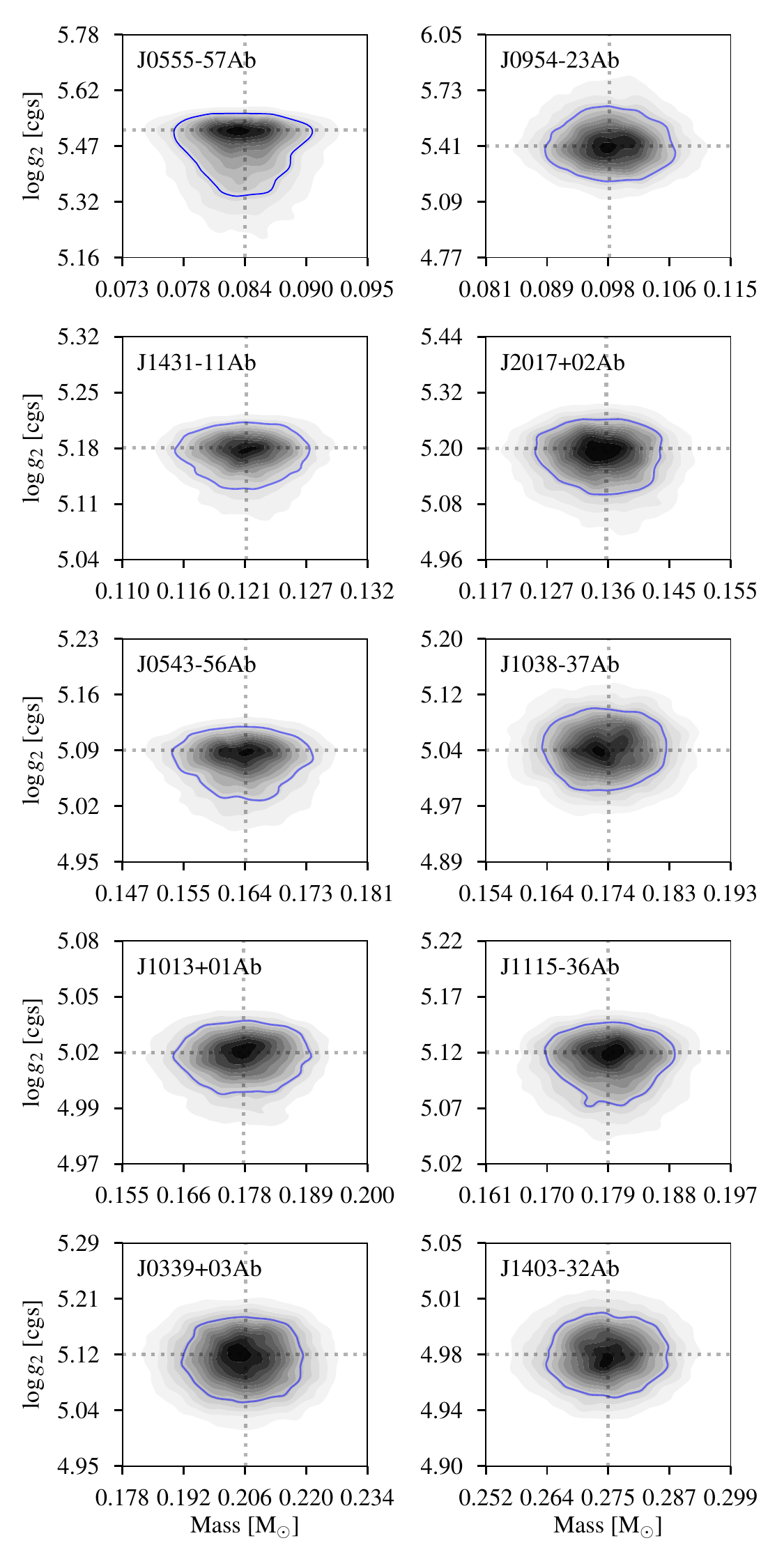}
        \caption{Mass-surface-gravity posterior distributions generated by the Markov chain Monte Carlo sampling. Visualised using kernel-density estimates. The 68$\%$-level contour is shown in blue. Modal values are indicated by dotted lines. The surface gravity can be determined without invoking the primary star mass, the companion mass and surface-gravity are uncorrelated.}
        \label{Mlogg_posteriors}
    \end{minipage}
\end{figure*}

\begin{figure*}
     \begin{minipage}[t][][t]{1\columnwidth}
        \centering
        \includegraphics[width=1\linewidth]{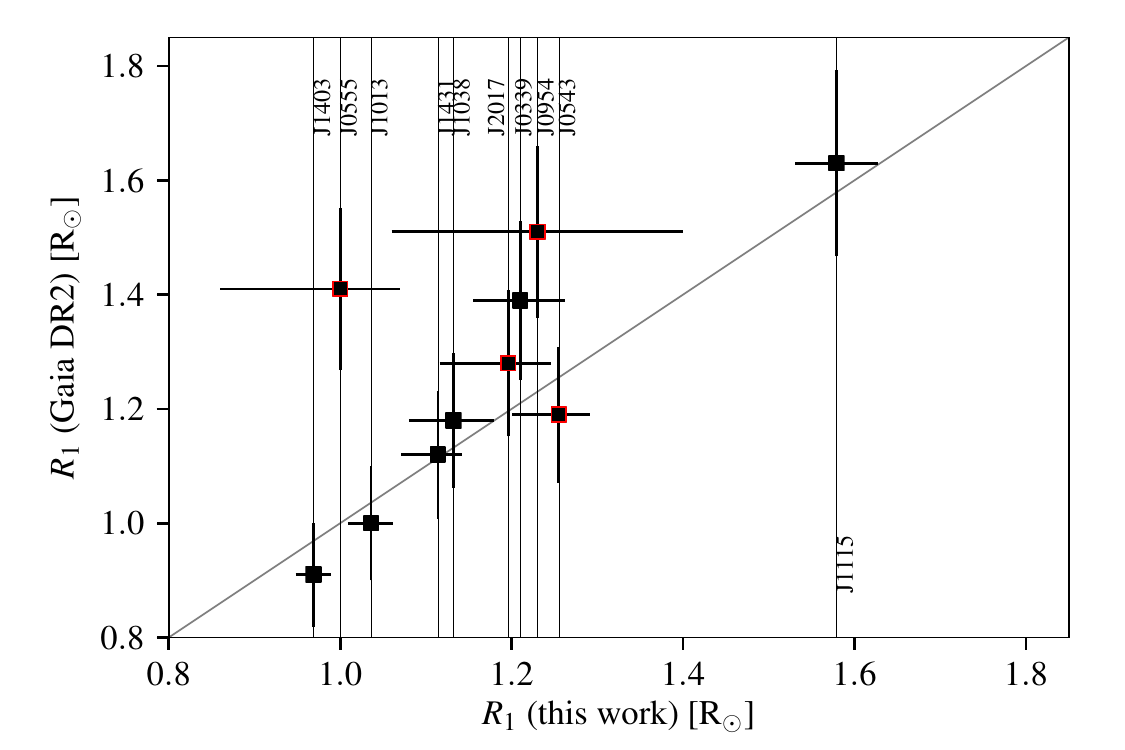}
	    \caption{Comparison of derived primary star radii with radius estimates from \textit{Gaia} DR2 three-band photometry and parallax measurements. Objects delineated in red have an astrometric goodness of fit parameter $\mathrm{GOF_{AL}} > 8$, indicating a poor fit of the \textit{Gaia} astrometric solution.}  
    	\label{fig:DR2-COMP}
    \end{minipage}
    \hfill 
    \begin{minipage}[t][][t]{1\columnwidth}
        \centering
    	\includegraphics[width=1\linewidth]{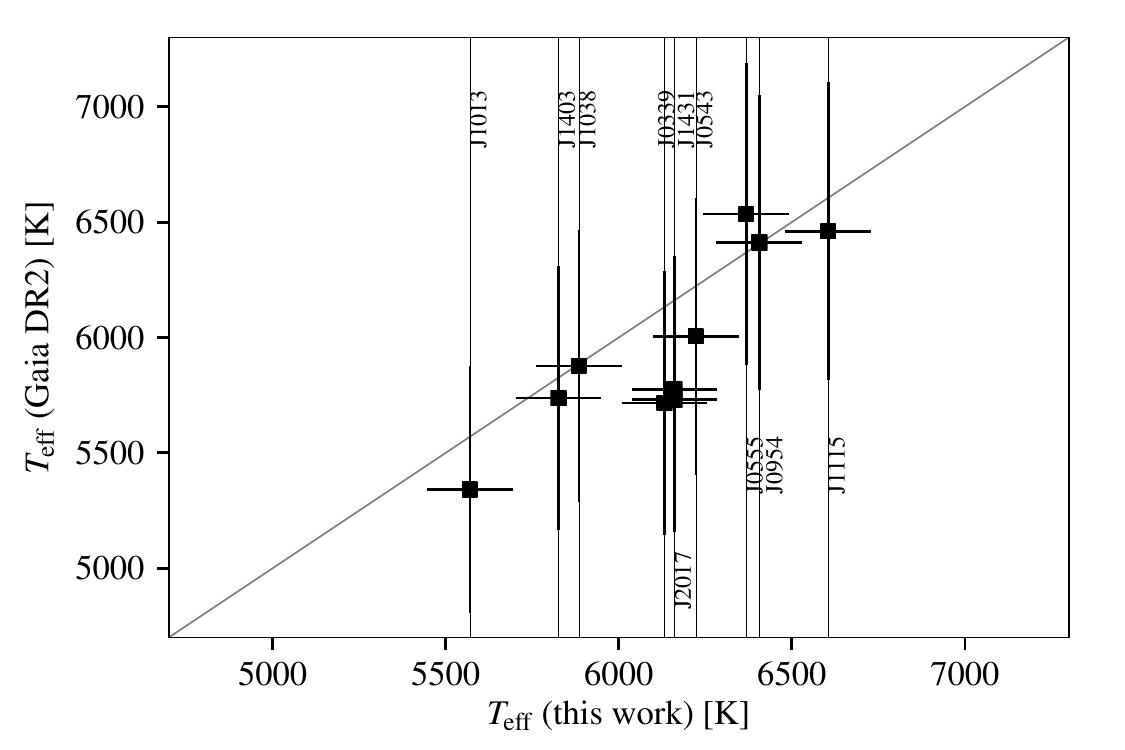}
	    \caption{Comparison of spectroscopic effective temperature measurements of the primary stars with temperature estimates from \textit{Gaia} DR2 three-band photometry.}  
	    \label{fig:DR2-COMP2}
    \end{minipage}
\end{figure*}

The thirteen added objects are very-low-mass stars orbiting heavier, bright host stars that permitted the spectroscopic measurement of the metallicity of each system. KOI-126C and KOI-126B \citep{Carter_2011} are part of a hierarchical triple, in which C and B form a close eclipsing binary, that eclipses the 1.3 M$_\odot$KOI-126A. The thirteen stars span a broad period range, with 7 objects that have sub--5 day periods, and 5 objects with periods greater than 5 days, including the wide binaries Kepler-16 and KOI-686, both with $P > 40$ days. The eccentricities are consistent with the period threshold for circularized orbits of $\sim$5 days. We re-examined the relationship between metallicity and radius residuals for the combined sample of 23 stars. Figure \ref{fig:radius_residuals-adjusted-all} shows the radius residuals determined for the combined sample, with respect to the Exeter/Lyon stellar evolution models. The linear fit of the radius residuals determined,
\begin{equation}
\left(\frac{\Delta R}{R}\right)_{\rm Adj} = 
\left(0.12^{+0.10}_{-0.06}\right)\rm{[Fe/H]} + \left(0.007^{+0.012}_{-0.011}\right).
\end{equation} 
A comparison with a one-parameter, constant model determined $\Delta_{\rm BIC} = 1.97 < 6$, suggesting that the radius residuals and metallicity are uncorrelated. 
The sample of literature objects includes stars with significantly super-solar metallicities, notably HATS-550-016, with [Fe/H] $= +0.60 \pm 0.06$ dex, and KIC1571511, with [Fe/H] $= +0.37 \pm 0.08$ dex. The Dartmouth super-solar isochrones plotted in Fig. ~\ref{fig:MASS-RADIUS} (dashed, grey) indicate that the scaling of metallicity in the super-solar regime is non-linear. Linearly extrapolating the sub-solar Exeter/Lyon isochrones then underestimates the effect of the metallicity, leading to overestimates of the radius residuals. In the mass range $M > 0.16$ M$_\odot$, where the Dartmouth and Exeter/Lyon isochrones are in good agreement, we used the Dartmouth isochrones to verify that the radius residuals are not significantly overestimated due to the extrapolation of the Exeter/Lyon models.
The objects in the literature sample lack a homogeneous method of analysis; this is particularly significant for the determination of the stellar metallicity. \citet{Jofre_2013} examined metallicity measurements for 34 F, G and K-type \textit{Gaia} benchmark stars, and determined that discrepancies of up to 0.5 dex are common, for metallicity measurements from different references, as shown in Figure 1 in \citet{Jofre_2013}. Sources of discrepancies include the adopted method of analysis, different spectral line lists, atmospheric models and solar abundances. Such inhomogeneities in the method of determining the stellar metallicity lead to an additional scatter, and may be sufficient to mask a correlation of radius residuals and metallicity. 

\subsection{Effect of short orbital periods in the extended sample}
The comparison of the radii of short-period ($P \lesssim 5$ days) stars with the radii of long-period ($P \gtrsim 5$ days) objects was repeated for the combined sample of 23 stars. The radius residuals are plotted against the orbital periods of the binary systems in Figure \ref{fig:radius_residuals-adjusted-all_period}. This determined mean radius residuals of $\mu_c = 0.010 \pm 0.007$ for sub-5 day period, circularised objects, and $\mu_e = 0.024 \pm 0.006$ for objects with $P > 5$ days, and eccentric orbits. Both samples indicate a weak systematic inflation. A linear fit of the residuals as a function of the orbital period determined no evidence for a correlation across the complete period range, $\Delta R/R = (-0.0002 \pm 0.0002)P - (0.012 \pm 0.004)$. Several short-period objects from the literature are consistent with model predictions (eg. KOI-126C and B), or have negative radius residuals (HATS-551-019), similar to the short period objects EBLM J2017+02Ab and EBLM J0339+03Ab. On the other hand, the long-period Kepler-16B ($P = 41$ days) is significantly inflated with respect to isochrone predictions. 

\section{Conclusion}

Eclipse observations and spectroscopic measurements of the radial velocities of ten binary stars were used to make measurements of stellar masses and radii in the fully convective very-low-mass regime, $M < 0.35$ M$_\odot$. The data were analysed in a fully Bayesian framework and significantly increase the number of well-characterised low-mass stars in the fully convective regime. The measurements have a mean precision in mass and radius of 4.2$\%$ and 7.5\% respectively; some objects have radii determined to a precision better than 5\%. The bright solar-like primary star in each binary system prevents spectroscopic measurements of the radial velocity of the low-mass companions, but permits a robust spectroscopic measurement of the stellar metallicity that is otherwise difficult to obtain for very-low-mass stars.

The Exeter/Lyon stellar evolution models were used to compare the empirical radii of the ten very-low-mass stars with theoretical predictions. The radius residuals indicate a marginal underestimate of stellar radii by the Exeter/Lyon models, with a mean radius residual of 1.6 $\pm 1.2 \%$. The extended sample of 23 fully convective stars determined a mean radius residual of 1.9 $\pm\ 0.5 \%$. No systematic radius inflation of short-period low-mass stars over their longer period counterparts could be reliably established. A correlation of radius residuals with metallicity in the sample of ten stars is not robust to omitting the significantly discrepant and precise radius measurement of EBLM J1013+01Ab, or to extending the sample to 23 stars. The radius of the short-period low-mass star EBLM J1013+01Ab exceeds theoretical predictions by $6.4^{+2.0}_{-1.9} \%$, after accounting for the metallicity of its host star.

The inflation of the radii of low-mass stars is frequently associated with magnetic effects \citep{Mullan_MacDonald_2001,Chabrier_2007} that are known to be particularly significant in rapidly rotating objects, such as tidally locked components in short-period binary systems \citep{Pizzolato_2003, Wright_2011}. In the fully convective regime such effects may be less significant, since a shell-type dynamo cannot operate. The consistency of the present sample of very-low-mass stars with the Exeter/Lyon stellar models appears to be better than that of the higher-mass sample of low-mass stars in binaries shown in Figure \ref{fig:MASS-RADIUS-all}. Significantly inflated radii in the fully convective regime are nevertheless established, here (EBLM J1013+01Ab), and elsewhere \citep[][]{Parsons_2018}.
Several fast-rotating objects are consistent with stellar models (eg. EBLM J2017+02Ab, KOI-126C\&B), suggesting that other stellar properties are effective at controlling the stellar radius. Our analysis indicates that the stellar metallicity may have a measurable effect on the stellar radius, but cannot be invoked to reconcile substantially inflated objects with the Exeter/Lyon stellar evolution models. To examine the effect of the metallicity on the stellar radius, a large sample of empirical radii and homogeneous metallicity measurements is desirable. The detection of hundreds of low-mass eclipsing binary stars as false positives in searches for transiting exoplanets \citep{EBLM_4,Collins_2018} will permit robust statistical studies to determine the physical properties of such objects. We expect that the number of measurements of masses and radii of very-low-mass stars will substantially increase as results from the recently launched TESS mission \citep{TESS} are becoming available. These measurements will enable precise tests of stellar evolution models that will refine our understanding of the structure of very-low-mass stars and of their exoplanets. 

\begin{acknowledgements}
  The authors thank the anonymous referee for valuable comments on the manuscript. SG acknowledges support from Doctoral Training Partnership grant number
  ST/N504348/1 from the Science and Technology Facilities Council (STFC). PM
  acknowledges support from STFC research grant number ST/M001040/1. 
  
  The research leading to these results has received funding from the European Research Council under the FP/2007-2013 ERC Grant Agreement n$^\circ$ 336480; the European Union’s Horizon 2020 research and innovation program, grant agreement n$^\circ$ 803193/BEBOP; and from the ARC grant for Concerted Research Actions, financed by the Wallonia-Brussels Federation. MG and EJ are F.R.S.-FNRS Senior Research Associates, and VVG is F.R.S-FNRS Research Associate.
  DQ and LD acknowledge support from the Simons Foundation (PI: Queloz, grant number 327127).
  
  The Swiss Euler Telescope is funded by the Swiss National Science Foundation. TRAPPIST-South is a project funded by the Belgian Fonds (National) de la Recherche Scientifique (F.R.S.-FNRS) under grant FRFC 2.5.594.09.F, with the participation of the Swiss National Science Foundation (FNS/SNSF). WASP-South is hosted by the South African Astronomical Observatory.
  
  This work has made use of data from the European Space Agency (ESA) mission {\it Gaia} (\url{https://www.cosmos.esa.int/gaia}), processed by the {\it Gaia} Data Processing and Analysis Consortium (DPAC, \url{https://www.cosmos.esa.int/web/gaia/dpac/consortium}). Funding for the DPAC has been provided by national institutions, in particular the institutions participating in the {\it Gaia} Multilateral Agreement.
\end{acknowledgements}

\bibliographystyle{aa}
\bibliography{references}

\begin{table*}[h!]
	\renewcommand{\arraystretch}{1.6}
	\centering
	\caption{Very-low-mass eclipsing binary stars selected from the literature, that are fully convective and have a spectroscopic measurement of the system metallicity.}
	\label{Literature-objects}
	\begin{tabular*}{\textwidth}{ll|rrrrrrl}
		\hline \hline
		& Low-mass star &$M$ [M$_\odot$]&$R$ [R$_\odot$] & $P$ [d]& [Fe/H] [dex]& $\sigma_{\rm [Fe/H]}$ & $e$ & Reference \\
		\hline 
		&EBLM J1219-39Ab  & 0.0911$^{+(18)}_{-(24)}$ & 0.1146$^{+(70)}_{-(50)}$ &  6.76 & $-0.23$ & 0.08 & 0.06 &  \citet{EBLM_1}\\
		&OGLE-TR-122  & 0.0920$^{+(90)}_{-(90)}$ & 0.120$^{+(24)}_{-(13)}$&  7.27 &  0.15 & 0.36 & 0.21 & \citet{Pont_2005} \\
		&C4780b         & 0.096$^{+(11)}_{-(11)}$ & 0.105$^{+(14)}_{-(16)}$ & 20.68  &  0.20  & 0.20 & 0.40&\citet{Tal-Or_2013} \\
		&KOI-686        & 0.0987$^{+(46)}_{-(46)}$   & 0.1226$^{+(30)}_{-(30)}$ & 52.50     &  0.02 & 0.12 & 0.56 & \citet{Diaz_2014} \\
		&HATS-550-016 & 0.1100$^{+(50)}_{-(60)}$   & 0.1467$^{+(30)}_{-(40)}$ &  2.05 &  0.60  & 0.06 & 0.00&\citet{Zhou_2013} \\
		&EBLM J0113+31 & 0.186$^{+(10)}_{-(10)}$   & 0.209$^{+(11)}_{-(11)}$ &  14.28&  $-$0.41  & 0.06 & 0.31& \citet{Chew_2014}\\
		&HATS-551-021   & 0.1320$^{+(140)}_{-(50)}$  & 0.1537$^{+(60)}_{-(80)}$ &  3.64 & $-0.40$  & 0.10  & 0.00& \citet{Zhou_2013}\\
		&KIC1571511   & 0.1410$^{+(50)}_{-(40)}$   & 0.1779$^{+(20)}_{-(20)}$ &  14.02   &  0.37 & 0.08 & 0.33&\citet{Ofir_2011} \\
		&HATS-551-019   & 0.170$^{+(10)}_{-(10)}$ & 0.180$^{+(10)}_{-(10)}$ &  4.69 & $-0.40$  & 0.10 & 0.00&\citet{Zhou_2013} \\
		&HATS-553-001   & 0.200$^{+(10)}_{-(20)}$ & 0.22$^{+(10)}_{-(10)}$ &  3.80 & $-0.10$  & 0.20 & 0.00&\citet{Zhou_2013} \\
		&Kepler-16B    & 0.20256$^{+(67)}_{-(67)}$ & 0.22620$^{+(60)}_{-(50)}$* & 41.08 & $-0.04$**  & 0.08 & 0.16&\citet{Doyle_2011} \\
		&KOI-126C       &0.2127$^{+(25)}_{-(25)}$   & 0.2321$^{+(20)}_{-(20)}$ &  1.35   &  0.15 & 0.08 & 0.02 &\citet{Carter_2011}\\
		&KOI-126B       & 0.2413$^{+(31)}_{-(31)}$   & 0.2542$^{+(20)}_{-(20)}$ &  1.35   &  0.15 & 0.08 & 0.02&\citet{Carter_2011} \\
	\end{tabular*}
	\vspace{0.3cm}
	\tablefoot{*The low uncertainties in the radius of Kepler-16B are possible due to the double-eclipsing configuration and the circumbinary planet in the Kepler-16 system. **Adopted from the re-analysis by \citet{Winn_2011}. KOI-126C and B are part of a hierarchical triple.}
\end{table*}

\begin{table*}
	\renewcommand{\arraystretch}{1.6}
	\centering
	\caption{Metallicity-adjusted fractional radius residuals for the sample of ten very-low-mass stars, and for 13 very-low-mass stars from the literature, with respect to the Exeter/Lyon models \citep{Baraffe_98,Baraffe_2003,Baraffe_2015}, and the Dartmouth models \citep{Dotter_2008} if $M_2 > 0.15$ M$_\odot$. References for the literature objects are provided in Table \ref{Literature-objects}.}
	\label{Radius-residuals}
	\begin{tabular*}{\textwidth}{l|rrrrrr}
		\hline \hline 
		Low-mass star & $\Delta R/ R$ [Baraffe] & $\sigma_+$ & $\sigma_{-}$ & $\Delta R / R$ [Dartmouth] & $\sigma_+$  &  $\sigma_{-}$  \\
		\hline 		
		\multicolumn{7}{l}{\textit{This work}} \\
		EBLM J0339+03Ab & $-0.056$ & 0.055  & 0.057  & $-0.045$ & 0.056 & 0.054 \\
        EBLM J0954-23Ab & $-0.139$ & 0.170  & 0.241  & -- & -- & -- \\
        EBLM J1013+01Ab &  0.064   & 0.020   & 0.019  & 0.076 & 0.023 & 0.022 \\
        EBLM J1431-11Ab &  0.001   & 0.042  & 0.031  & -- & -- & -- \\
        EBLM J1403-32Ab & $-0.010$ & 0.025  & 0.025  & 0.008 & 0.026 & 0.028 \\
        EBLM J1038-37Ab &  0.037   & 0.044  & 0.046  & 0.054 & 0.047 & 0.047 \\
        EBLM J2017+02Ab & $-0.035$ & 0.077  & 0.066  & -- & -- & -- \\
        EBLM J0543-56Ab &  0.019   & 0.046  & 0.031  & 0.038 & 0.045 & 0.036 \\
        EBLM J1115-36Ab & $-0.040$ & 0.037  & 0.030  & $-0.022$ & 0.038 & 0.029 \\
        EBLM J0555-57Ab & $-0.218$ & 0.148  & 0.100  & -- & -- & -- \\
		\multicolumn{7}{l}{\textit{Literature objects}} \\
		EBLM J1219-39  &  0.039   & 0.055  & 0.059  & -- & -- & -- \\
        OGLE-TR-122Ab  &  0.044   & 0.167  & 0.204  & -- & -- & -- \\
        C4780b         & $-0.133$ & 0.202  & 0.254  & -- & -- & -- \\
        KOI-686Ab      & $-0.013$ & 0.053  & 0.053  & -- & -- & -- \\
        HATS-550-016Ab &  0.041   & 0.047  & 0.047  & -- & -- & -- \\
        HATS-551-021Ab &  0.007   & 0.074  & 0.077  & -- & -- & -- \\
        KIC1571511Ab   &  0.036   & 0.027  & 0.027  & -- & -- & -- \\
        HATS-551-019Ab & $-0.038$ & 0.071  & 0.079  & $-0.033$ & 0.075 & 0.078 \\
        HATS-553-001Ab &  0.012   & 0.073  & 0.078  & 0.031 & 0.072 & 0.077 \\
        Kepler-16B     &  0.030   & 0.006  & 0.006  & 0.031 & 0.006 & 0.006 \\
        KOI-126C       &  0.003   & 0.012  & 0.012  & 0.005 & 0.013 & 0.013 \\
        KOI-126B       & $-0.002$ & 0.012  & 0.012  & $-0.002$ & 0.013 & 0.013 \\
        EBLM J0113+31Ab&  0.041   & 0.061  & 0.068  & 0.039 & 0.064 & 0.068 \\
	\end{tabular*}
	\vspace{0.2cm}
	\label{Residual_table}
\end{table*}

%%%%% APPENDIX %%%%%
\begin{appendix}
\section{Radial velocity and eclipse measurements}
\begin{figure} 
	\centering
	\includegraphics[width=\columnwidth]{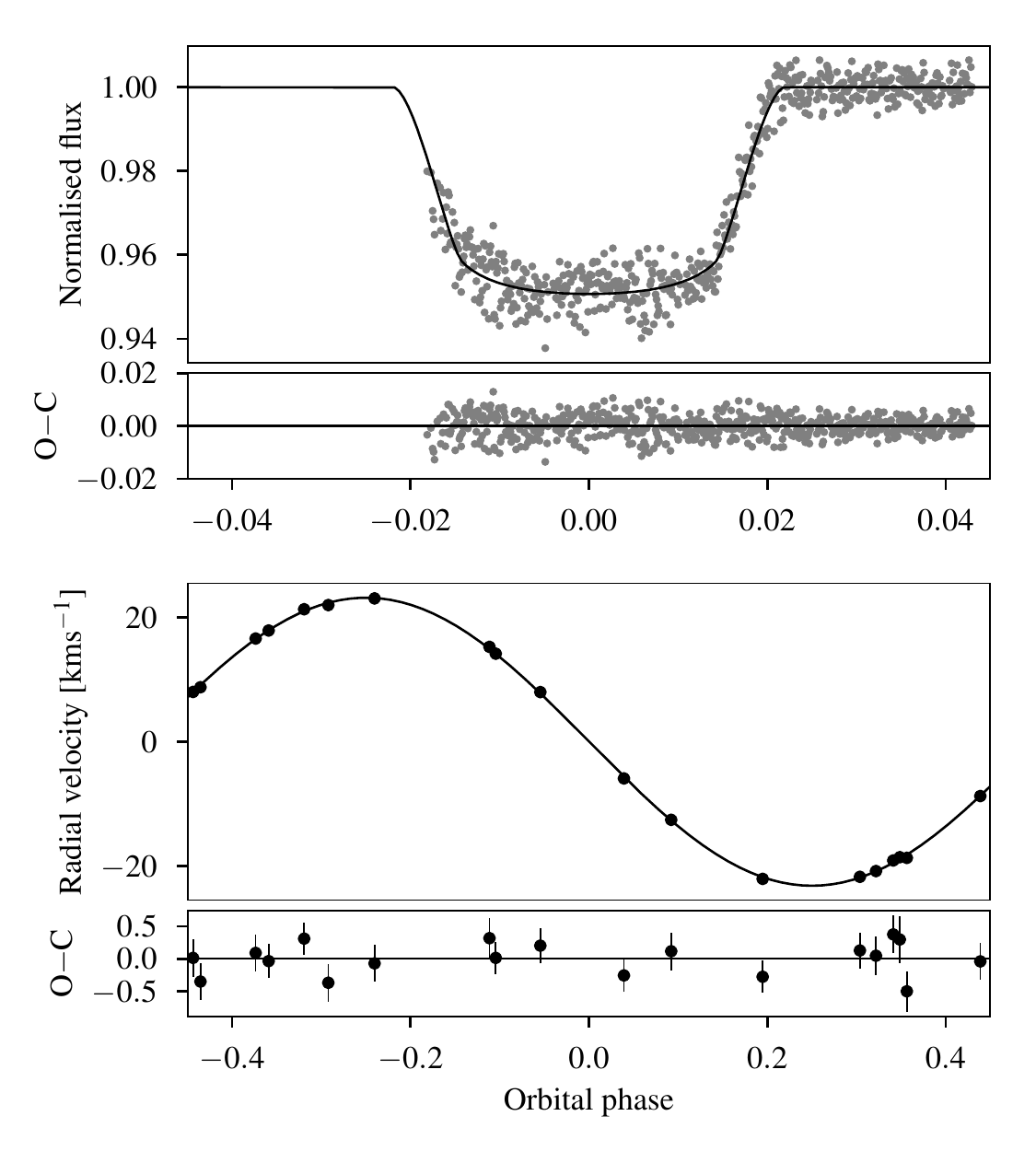}
	\caption{Top: Transit of EBLM J1013+01Ab (TRAPPIST), with model. Residuals (O-C) are shown in the lower panel. Lower: Radial velocity measurements for EBLM J1013+01A (CORALIE), with model. Residuals (O-C) are shown in the lower panel.} 
	\label{J1013_plots}
\end{figure}
\begin{figure} 
	\centering
	\includegraphics[width=\columnwidth]{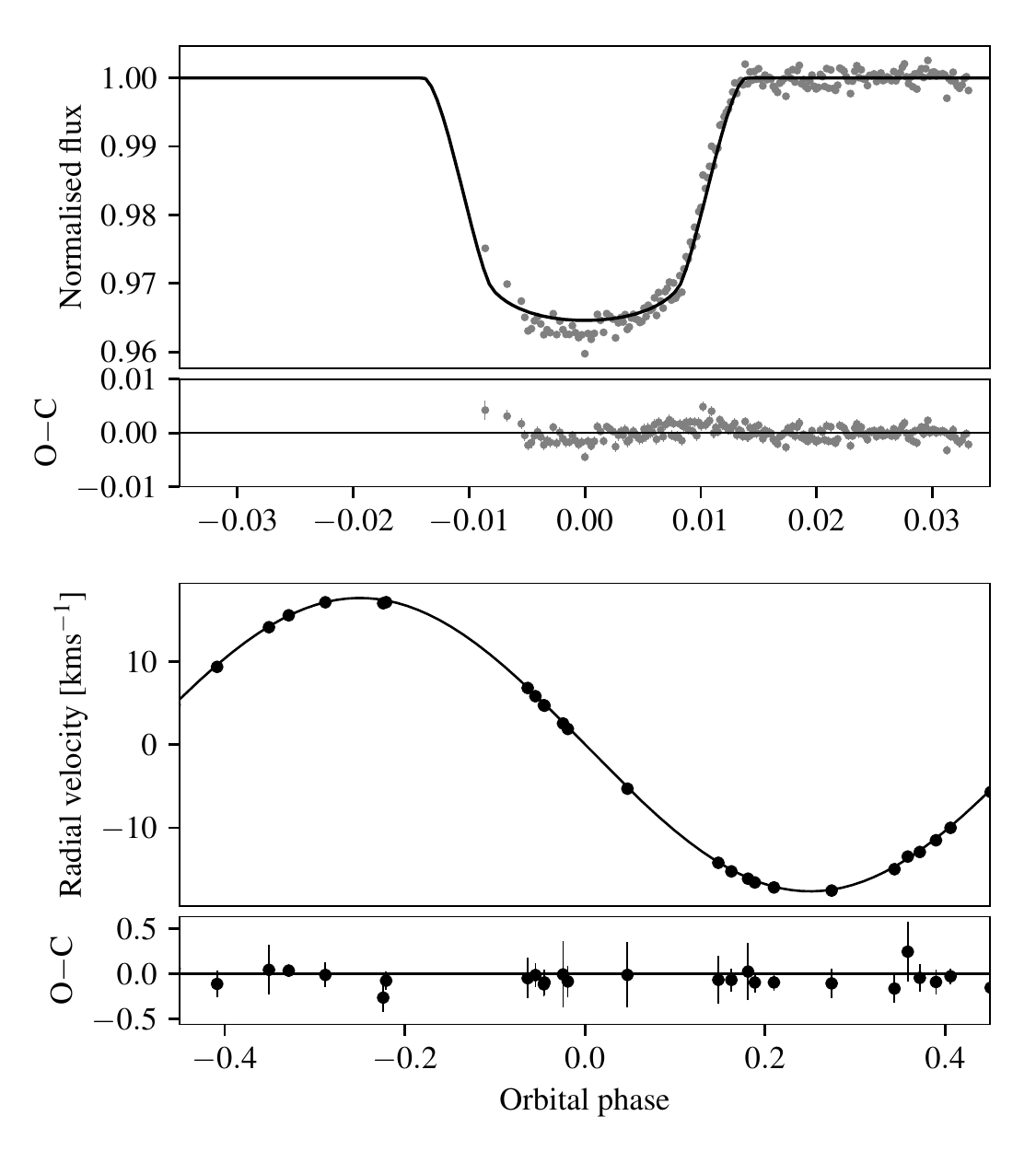}
	\caption{Top: Transit of EBLM J1038-37Ab  (Euler), with model. Residuals (O-C) are shown in the lower panel. Lower: Radial velocity measurements for EBLM J1038-37A (CORALIE), with model. Residuals (O-C) are shown in the lower panel.} 
	\label{J1038_plots}
\end{figure}
\begin{figure} 
	\centering
	\includegraphics[width=\columnwidth]{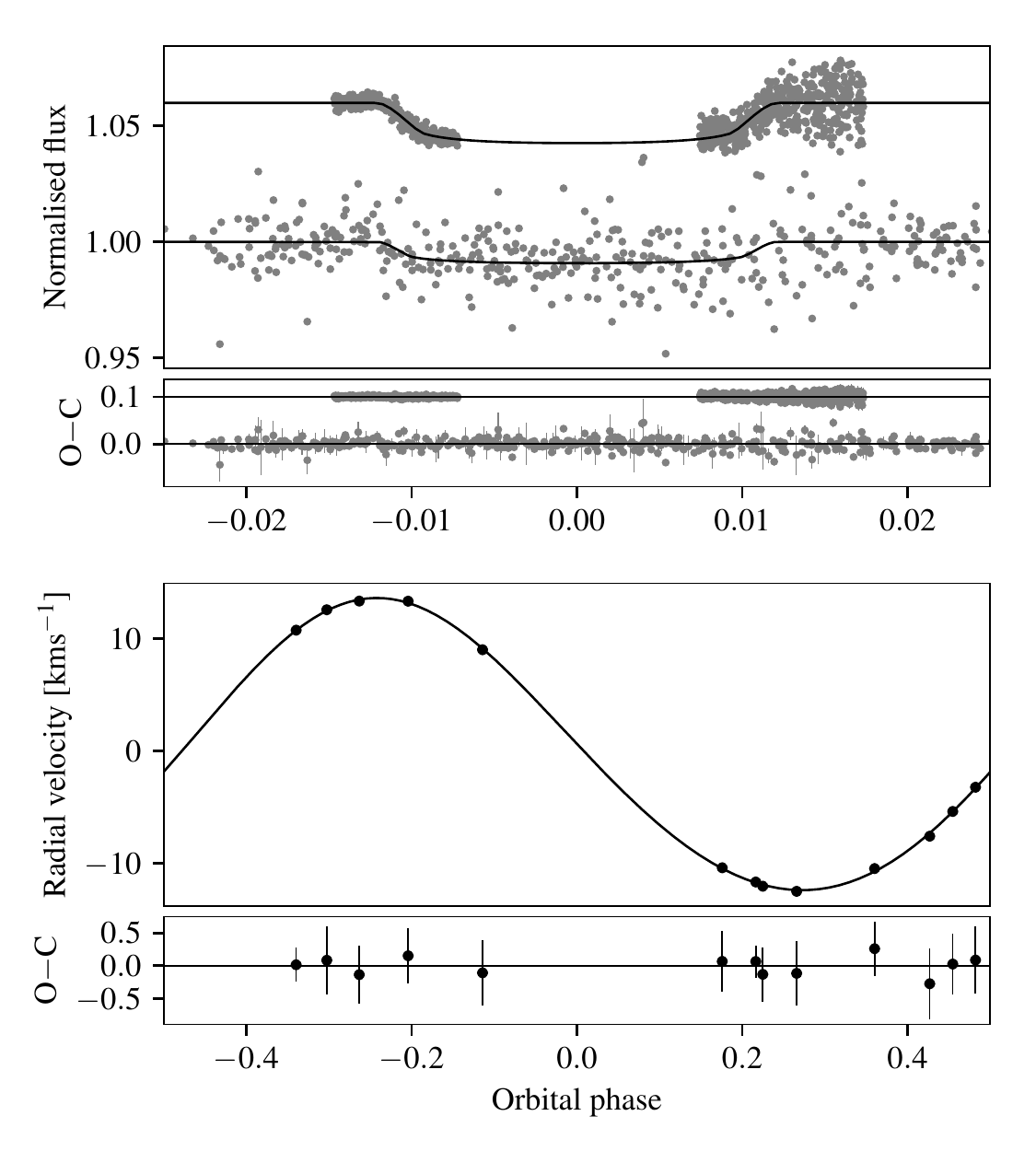}
	\caption{Top: Transit of EBLM J1115-36Ab with WASP (top data), SPECULOOS (ingress), and TRAPPIST (egress). The transit depth for WASP was fitted independently. Lower: Radial velocity for EBLM J1115-36A (CORALIE). Residuals (O-C) are shown in the lower panel.} 
	\label{J1115_plots}
\end{figure}
\begin{figure} 
	\centering
	\includegraphics[width=\columnwidth]{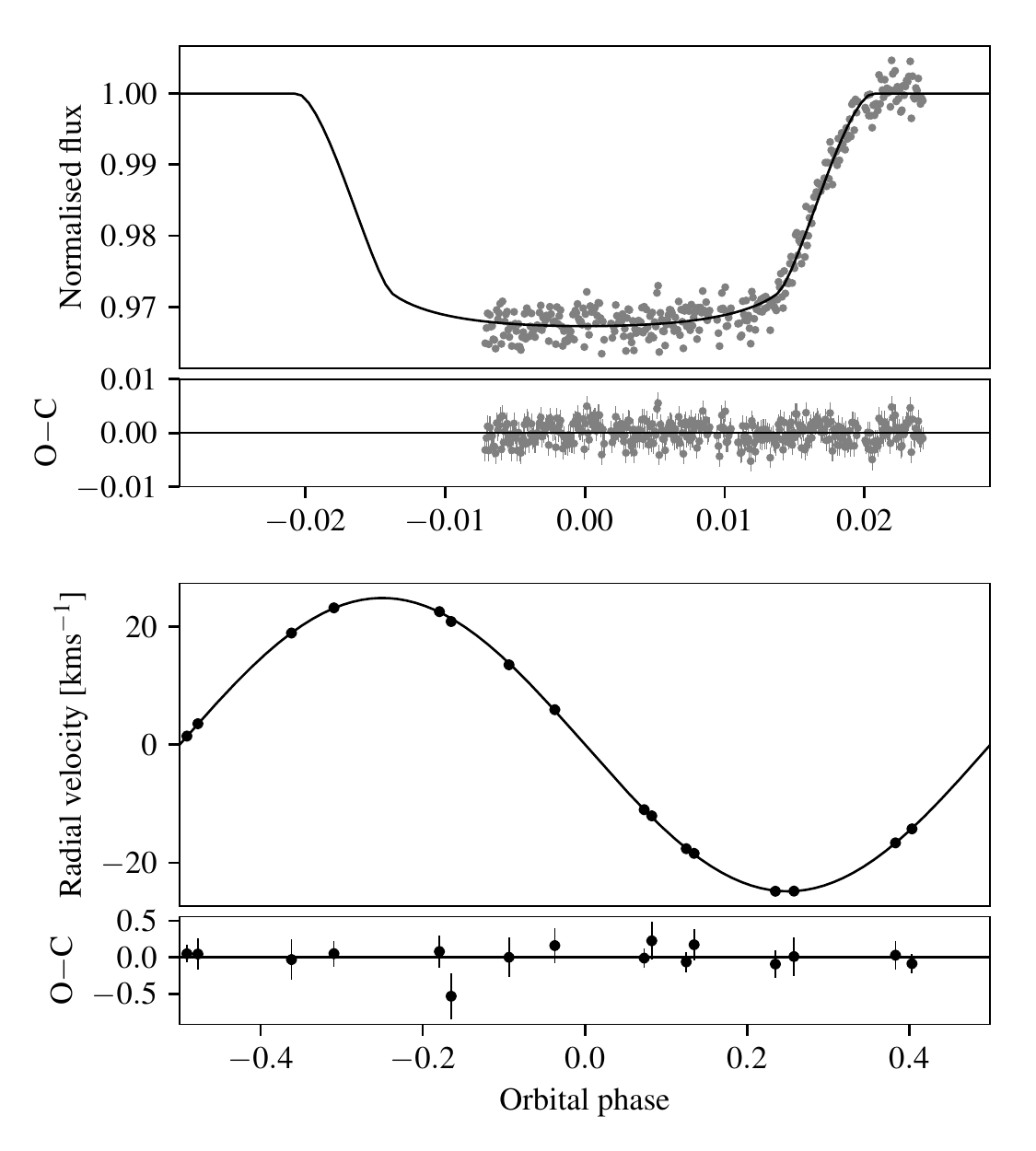}
	\caption{Top: Transit of EBLM J0339+03Ab (TRAPPIST), with model. Residuals (O-C) are shown in the lower panel. Lower: Radial velocity measurements for EBLM J0339+03A (CORALIE), with model. Residuals (O-C) are shown in the lower panel.} 
	\label{J0339_plots}
\end{figure}
\begin{figure} 
	\centering
	\includegraphics[width=\columnwidth]{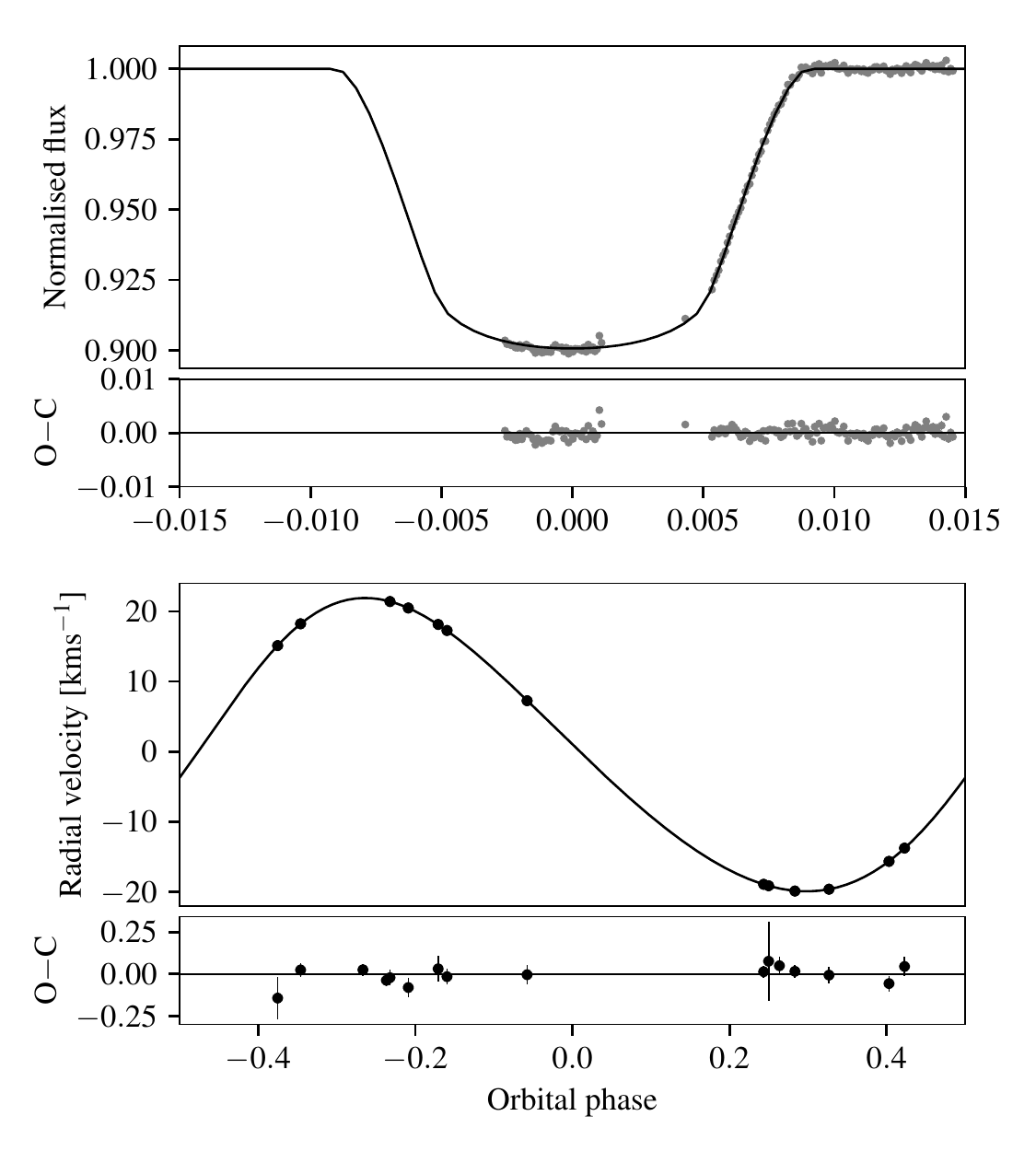}
	\caption{Top: Transit of EBLM J1403-32Ab (TRAPPIST), with model. Residuals (O-C) are shown in the lower panel. Lower: Radial velocity measurements for EBLM J1403-32A (CORALIE), with model. Residuals (O-C) are shown in the lower panel.} 
	\label{J1403_plots}
\end{figure}
\begin{figure} 
	\centering
	\includegraphics[width=\columnwidth]{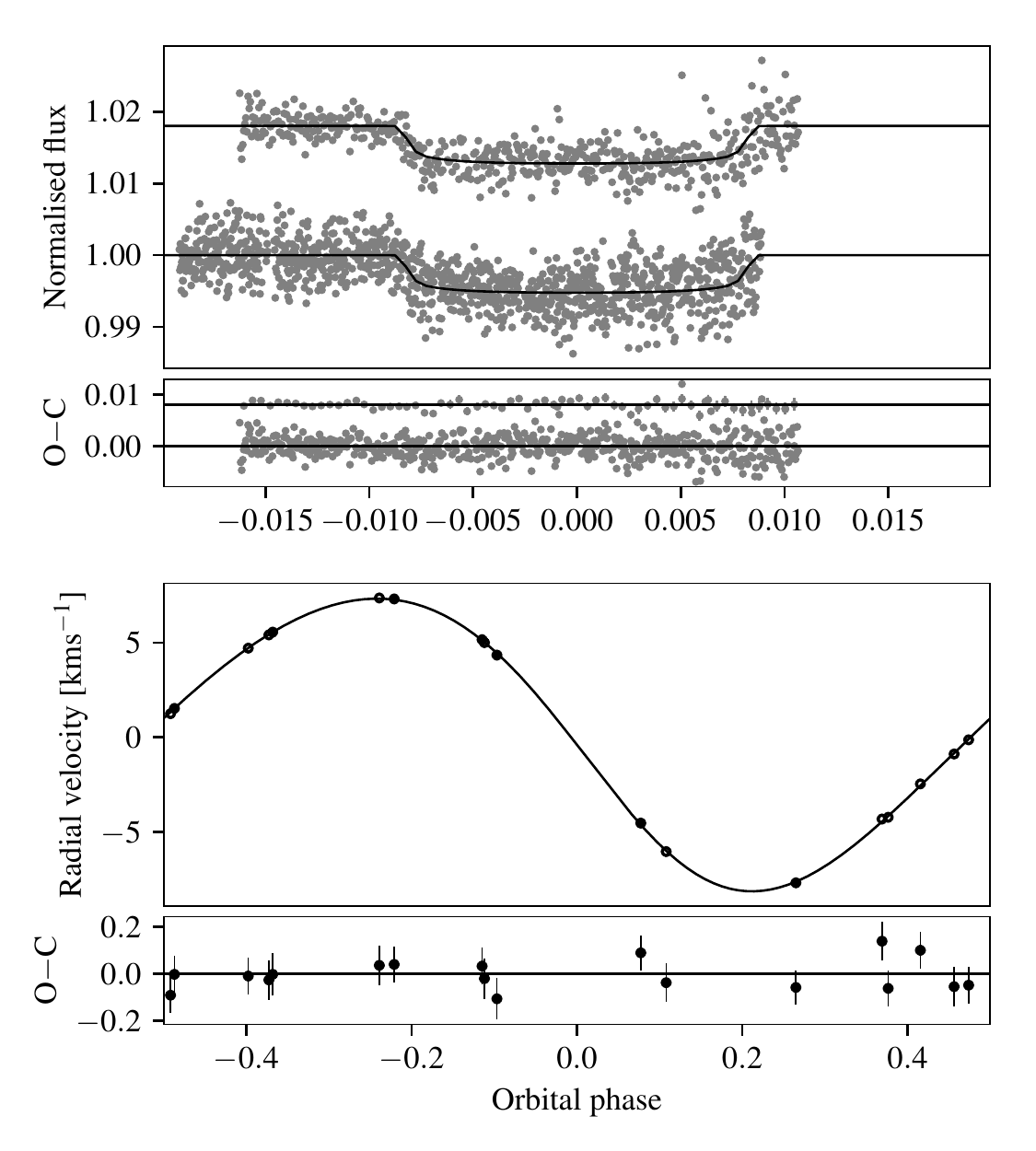}
	\caption{Top: Transits of EBLM J0555-57Ab (Euler, TRAPPIST), with model. Residuals (O-C) are shown in the lower panel. Lower: Radial velocity measurements for EBLM J0555-57A (CORALIE), with model. Residuals (O-C) are shown in the lower panel.} 
	\label{J0555_plots}
\end{figure}
\begin{figure} 
	\centering
	\includegraphics[width=\columnwidth]{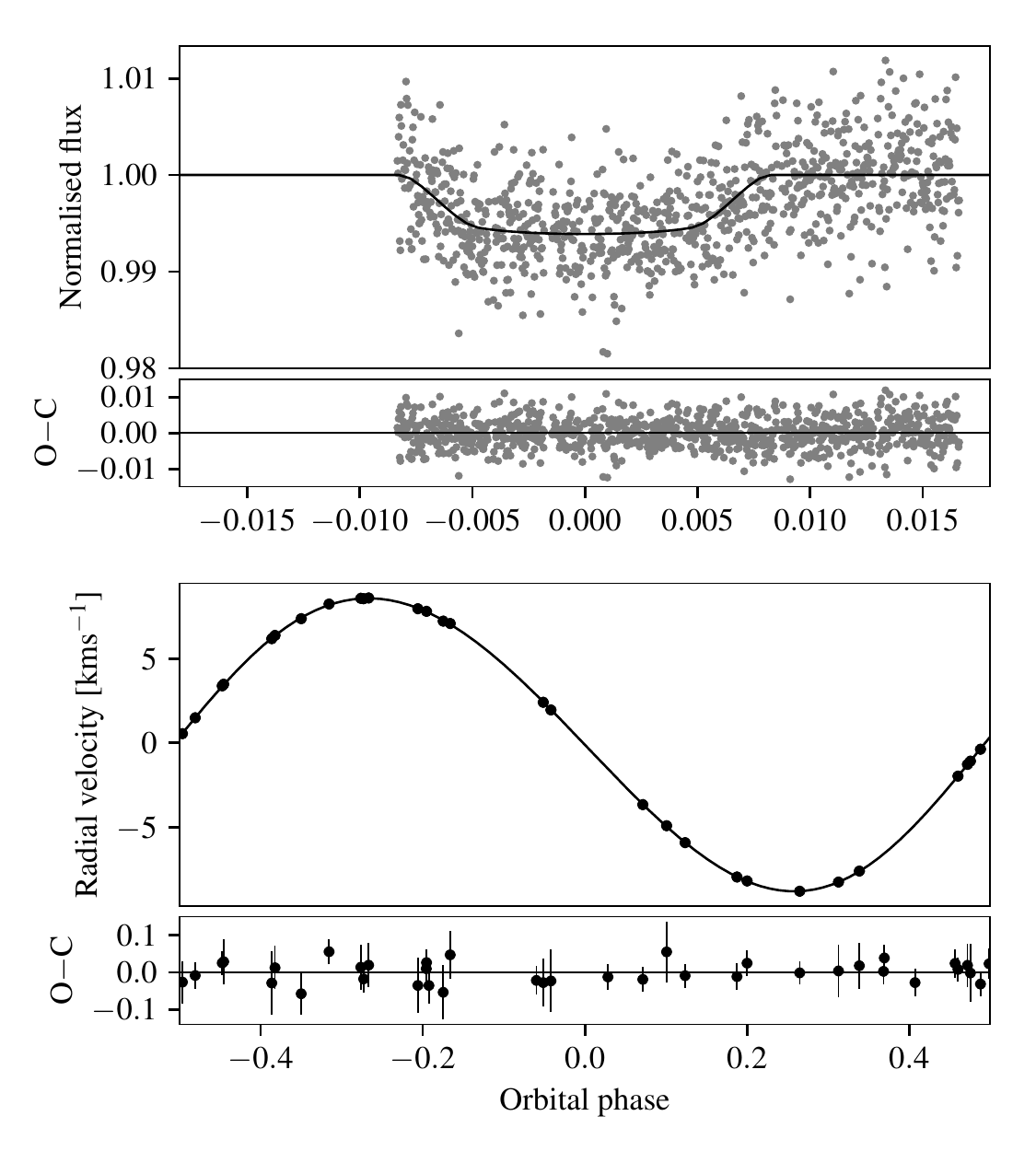}
	\caption{Top: Transit of EBLM J0954-23Ab (TRAPPIST), with model. Residuals (O-C) are shown in the lower panel. Lower: Radial velocity measurements for EBLM J0954-23A (CORALIE), with model. Residuals (O-C) are shown in the lower panel.} 
	\label{J0954_plots}
\end{figure}
\begin{figure} 
	\centering
	\includegraphics[width=\columnwidth]{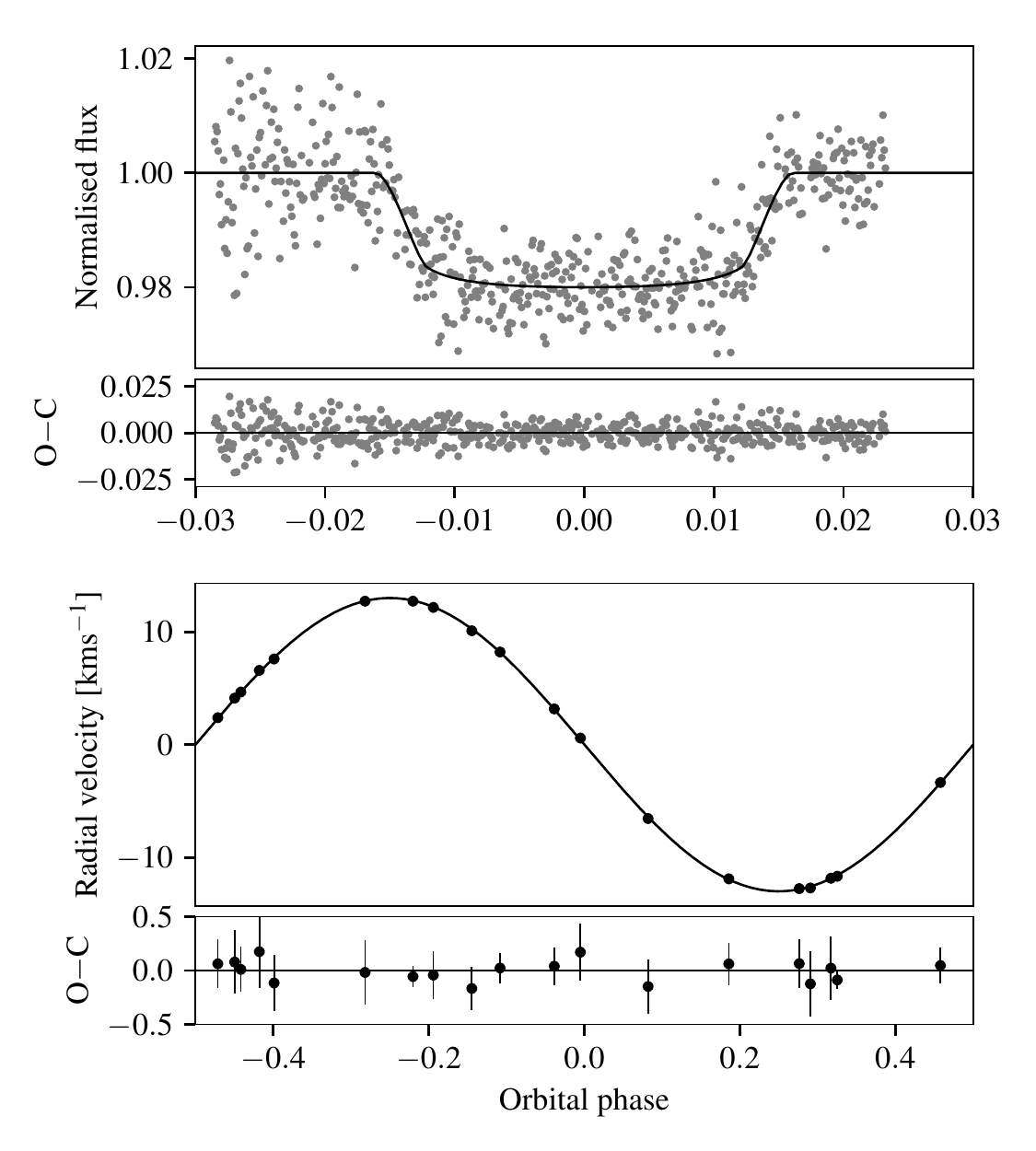}
	\caption{Top: Transit of EBLM J1431-11Ab (TRAPPIST), with model. Residuals (O-C) are shown in the lower panel. Lower: Radial velocity measurements for EBLM J1431-11A (CORALIE), with model. Residuals (O-C) are shown in the lower panel.} 
	\label{J1431_plots}
\end{figure}

\begin{figure*}
     \begin{minipage}[t][][t]{1\columnwidth}
        \centering
        \includegraphics[width=\linewidth]{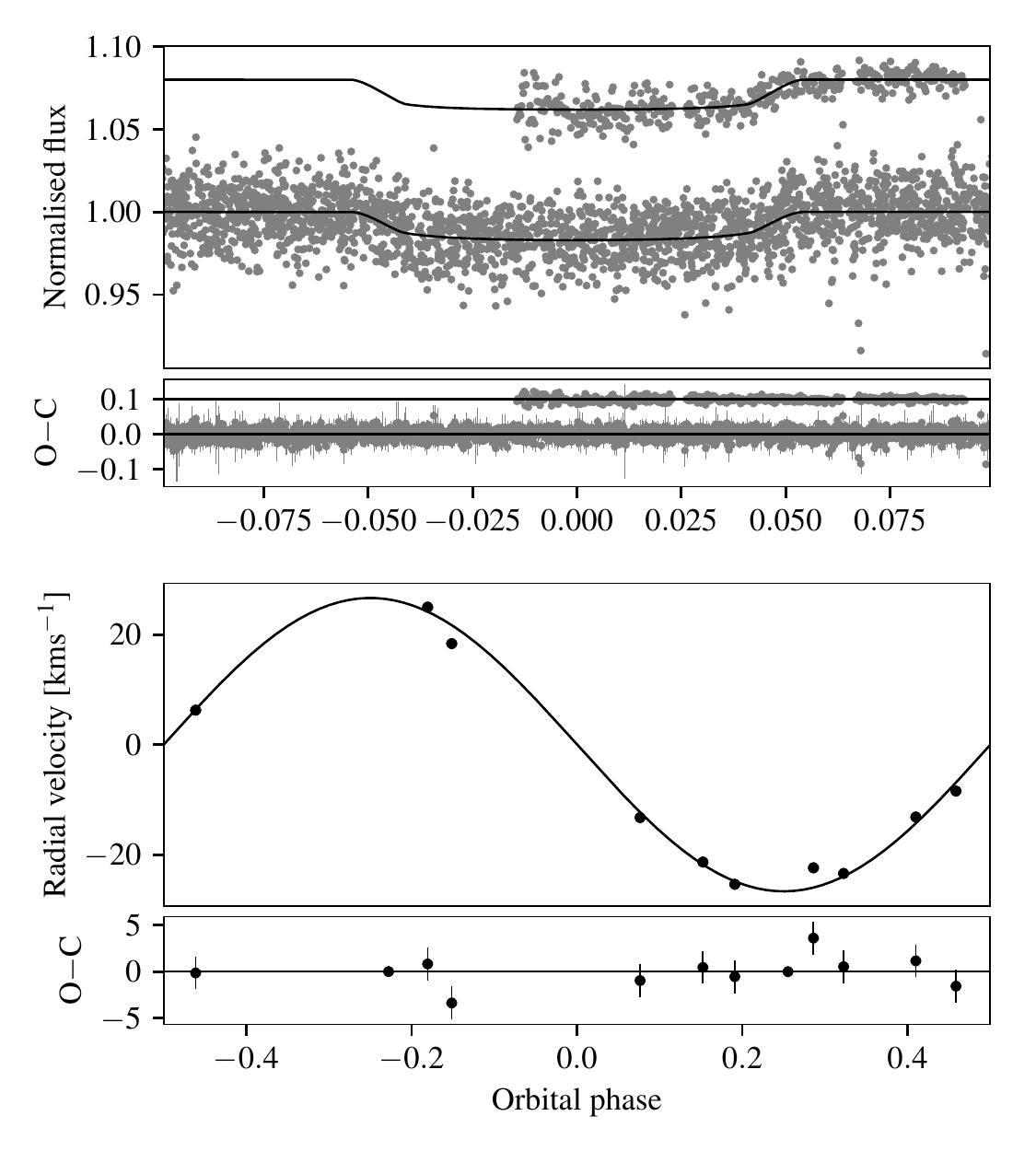}
	    \caption{Top: Transit of EBLM J2017+02Ab (TRAPPIST, egress) and WASP (lower data). Residuals (O-C) are shown in the lower panel. Lower: Radial velocity measurements for EBLM J2017+02A (CORALIE), with model. Residuals (O-C) are shown in the lower panel.} 
	\label{J2017_plots}
    \end{minipage}
    \hfill 
    \begin{minipage}[t][][t]{1\columnwidth}
        \centering
    	\includegraphics[width=\linewidth]{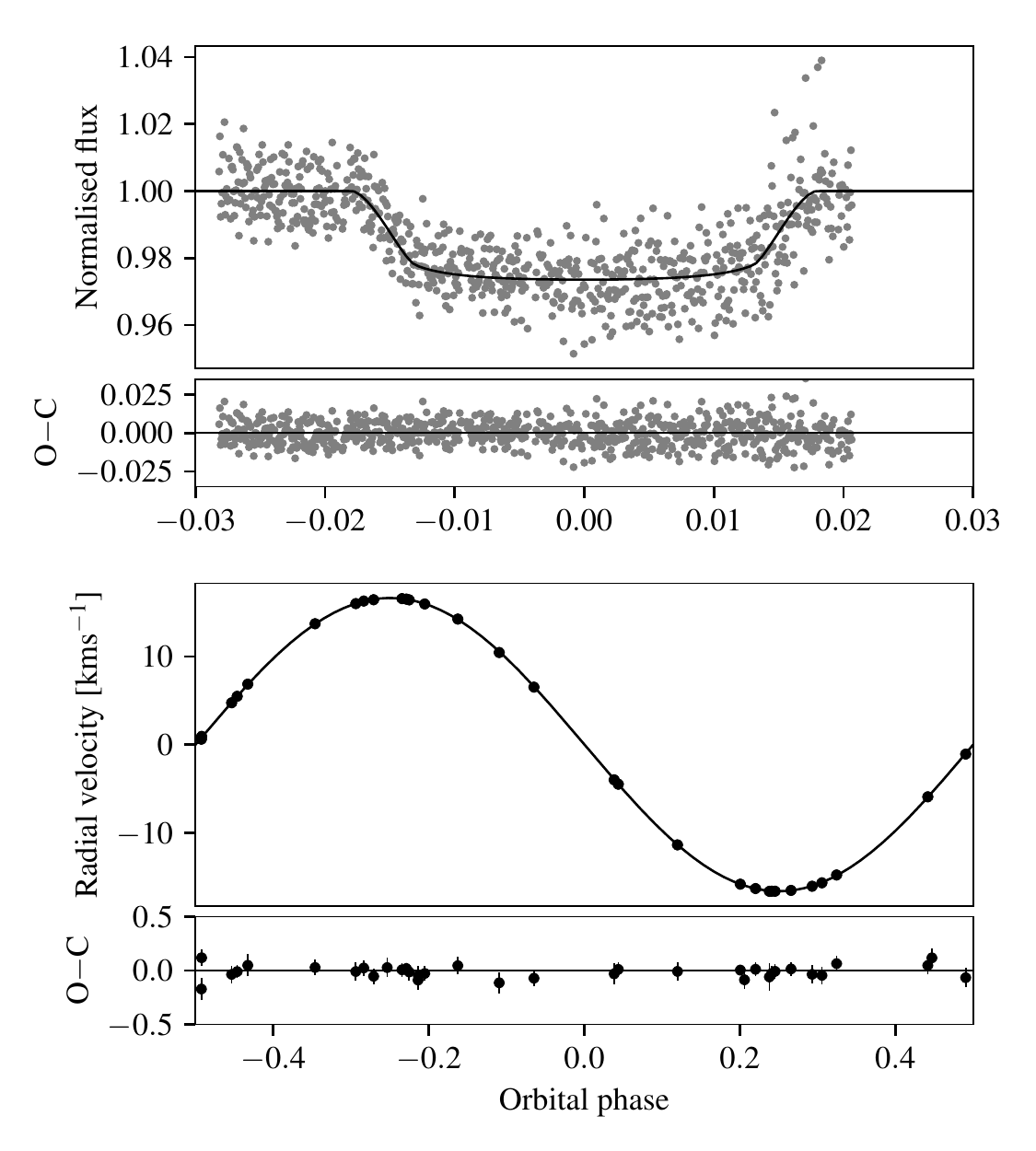}
	    \caption{Top: Transit of EBLM J0543-36Ab (TRAPPIST). Residuals (O-C) are shown in the lower panel. Lower: Radial velocity measurements for EBLM J0543-36A (CORALIE, HARPS), with model. Residuals (O-C) are shown in the lower panel.} 
	\label{J0543_plots}
    \end{minipage}
\end{figure*}

\end{appendix}
\end{document}